\newif\ifreport\reporttrue

\documentclass[conference]{IEEEtran}

\usepackage{cite}

\ifCLASSINFOpdf
   \usepackage[pdftex]{graphicx}
\else
\fi
\usepackage[cmex10]{amsmath}

\usepackage{amssymb, amsmath, amsfonts,dsfont}

\usepackage{algorithmic}

\usepackage{array}

\usepackage{mdwmath}
\usepackage{mdwtab}

\usepackage{eqparbox}

\usepackage[tight,footnotesize]{subfigure}
\usepackage{epsfig}

\usepackage{fixltx2e}

\usepackage{stfloats}

\newtheorem{lem}{Lemma}
\newtheorem{rem}{Remark}

\newtheorem{thm}{Theorem}
\newtheorem{defn}{Definition}

\begin{document}
%
\title{Achieving Delay Rate-function Optimality in OFDM Downlink with Time-correlated Channels}



%
\author{
\IEEEauthorblockN{Zhenzhi Qian\IEEEauthorrefmark{1},
Bo Ji\IEEEauthorrefmark{2},
Kannan Srinivasan\IEEEauthorrefmark{1}, 
Ness B. Shroff\IEEEauthorrefmark{1}\IEEEauthorrefmark{3}}
\IEEEauthorblockA{\IEEEauthorrefmark{1}Department of Computer Science and Engineering, The Ohio State University, Columbus 43210, OH}
\IEEEauthorblockA{\IEEEauthorrefmark{2}Department of Computer and Information Sciences, Temple University, Philadelphia 19122, PA}
\IEEEauthorblockA{\IEEEauthorrefmark{3}Department of Electrical and Computer Engineering, The Ohio State University, Columbus 43210, OH}}


\maketitle

\begin{abstract}
There have been recent attempts to develop scheduling schemes  for downlink transmission in a single cell of a multi-channel (e.g., OFDM-based) cellular network. These works have been quite promising in that they have developed low-complexity index scheduling policies that are delay-optimal (in a large deviation rate-function sense). However, these policies require that the channel is ON or OFF in each time-slot with a fixed probability (i.e., there is no memory in the system), while the reality is that due to channel fading and doppler shift, channels are often time-correlated in these cellular systems. Thus, an important open question is whether one can find simple index scheduling policies that are delay-optimal even when the channels are time-correlated. In this paper, we attempt to answer this question for time-correlated ON/OFF channels. In particular, we show that the class of \textit{oldest packets first} (OPF) policies that give a higher priority to packets with a large delay is delay rate-function optimal under two conditions: 1) The channel is \textit{non-negatively correlated}, and 2) The distribution of the OFF period is \textit{geometric}. 
We use simulations to further elucidate the theoretical results. 
\end{abstract}



%
\IEEEpeerreviewmaketitle

\section{Introduction}
Orthogonal frequency division multiplexing (OFDM) is a digital multi-carrier modulation method that has been widely used in wideband digital communications. 
A practical and important application is the downlink phase of a single cell of OFDM-based cellular networks, 
where the wideband can be divided into a large number of orthogonal sub-carriers, which can be used to carry data for different users. 
In this system, the Base Station (BS) maintains a separate queue to store data packets requested by each user. When the sub-carrier seen by a user is in good channel condition, 
the sub-carrier can successfully transmit a packet to the user from its designated queue. 
We will focus on the setting of a single-hop multi-user multi-channel system and study the delay performance of this system from a large-deviations perspective. 

In wireless networks, a key problem that has been extensively studied is the design of high-performance scheduling policies. 
It is well known from the seminal work \cite{Tassiulas} that the MaxWeight policy is throughput-optimal, in the sense that it can stabilize the system under any feasible arrival rates. 
However, it has been shown in \cite{Small_buffer} that the MaxWeight policy sacrifices the delay performance (and may lead to very large queue lengths) for better throughput. 
This fact has motivated researchers to look for policies that can improve the delay performance measured by a queue-length-based metric. 
In \cite{Average_delay}, the authors showed that the maximum-throughput and load-balancing (MTLB) policy can achieve delay optimality for two special cases of  ON/OFF channels with a two-user system or a system that allows fractional server allocation. 
However, this problem becomes much harder in general cases. On the other hand, in cellular networks, minimizing average delay may cause a large delay for certain users that have stringent delay requirements.

Another line of works focus on designing scheduling policies that maximize the rate-function of the steady-state probability that the largest queue length exceeds a given threshold when the number of channels and users both go to infinity. 
In \cite{Rate_queue1} and \cite{Rate_queue2} the authors showed that their proposed policy can achieve both throughput optimality and queue length rate-function optimality. 
However, simulations in \cite{Lin_ITA} - \cite{Simulation} show that good queue length performance does not necessarily imply good delay performance.
In fact, queue-length-based policies usually suffer from the so called ``last packet'' problem, which occurs in the situation where a certain queue has a very small number of packets. Hence, this queue is rarely scheduled by the queue-length-based policies, resulting in large packet delays. 

To that end, a delay-based metric has been investigated in recent works in \cite{Lin_Tech_Report, bo_hybrid} and \cite{bo_greedy}. The authors developed several policies that achieve both throughput optimality and delay rate-function optimality (or near-optimality). Although the results hold for general arrivals (e.g., time-correlated arrivals are allowed), the channels are assumed to be \emph{i.i.d.} over time.
In practice, 
the current channel condition could depend on past channel conditions. Therefore, the following important question remains: \emph{How do we design a low-complexity scheduling policy that achieves provably good throughput and delay performance in the OFDM downlink system with time-correlated channels?}

While it is relatively straightforward to develop throughput optimal policies even for time-correlated channels, developing policies that are delay-optimal or delay-efficient for time-correlated channels remains an open problem. 

To that end, we are motivated to consider the following question: 
\emph{Can we find index scheduling policies that are delay-optimal even when the channels are time-correlated?}
In this paper, we provide a positive answer in some cases. Specifically, we analyze the delay rate-function of the class of \textit{oldest packets first} (OPF) policies which give a higher priority to packets with a large delay and present two conditions under which delay rate-function optimality can be achieved by any OPF policy.

The key contributions of this paper are summarized as follows. We use an alternating renewal process to model a general ON/OFF time-correlated channel. 
We first prove an upper bound on the delay rate-function for any scheduling policy.
Then, we analyze the delay rate-function of the class of OPF policies, which give a higher priority to older packets. 
We present two conditions and show that if both conditions are satisfied, delay rate-function optimality can be achieved by any OPF policy.
The first condition 
requires that the channel condition is \textit{non-negatively correlated} over time. This is often observed in practical time-correlated channels. The second condition requires that the ``OFF'' period distribution has the memoryless property, whereas the ``ON'' period distribution could be arbitrary. 

The rest of the paper is organized as follows. In Section \ref{Model}, we describe the system model and the performance metric. In Section \ref{Upper_bound}, we derive an upper bound on the rate-function for any possible policy, and in Section \ref{Analysis}, we obtain an achievable rate-function of the class of OPF policies. Then in Section \ref{Relationship}, we propose two conditions that imply delay rate-function optimality of the class of OPF policies. We conduct simulations to validate our theoretical results  in Section \ref{Simulation} and make concluding remarks in Section \ref{Conclusion}. 

\section{System Model}\label{Model}
We use a time-slotted multi-queue multi-server system to model the downlink phase of a single cell OFDM system. In particular, we assume that there are $n$ servers which stand for frequency sub-carriers. Furthermore, we assume the number of users is equal to the number of channels for ease of presentation \cite{bo_hybrid}. The Base Station maintains a queue/buffer to store packets requested by each user, hence there are also $n$ queues in the queueing system. (We use terms ``server'' and  ``channel'', ``queue'' and ``user'' interchangeably throughout this paper.) Next, we present several notations that will be used later in this paper. We use $Q_i$ to denote the queue associated to the $i$-th user, and use $S_j$ to denote the $j$-th server for $1\leq i,j\leq n$. We use $Q_i(t)$ to denote the queue length of queue $Q_i$ at the beginning of time-slot $t$ immediately after new packet arrivals. All queues are assumed to have infinite buffer size. Further, we use $W_i(t)$ to denote the head-of-line (HOL) delay of queue $Q_i$ at the beginning of time-slot $t$ and use $W(t)=\max_{1\leq i\leq n} W_i(t)$ to denote the largest packet delay in the system at the beginning of time-slot $t$. Finally, we use $\mathds{1}_A$ to denote the indicator function that indicates whether event $A$ occurs or not.

\begin{figure}[htbp]
  \centering
  \includegraphics[width=0.25\textwidth]{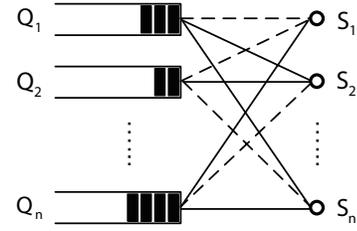}
  \caption{A multi-queue multi-server system with stochastic connectivity. The connectivity between queue $Q_i$ and server $S_j$ is ``ON'' if they are connected by a solid line, and ``OFF" otherwise (connected by a dashed line).}
\end{figure}
\subsection{Arrival Process}
The arrival process to each queue is assumed to be stationary and ergodic. We also assume the arrivals are \emph{i.i.d.} across all users, but could be correlated over time. Let $A_i(t)$ denote the number of packet arrivals to queue $Q_i$ in time-slot $t$. Let $A(t)=\sum_{i=1}^{n}A_i(t)$ denote the total packet arrivals coming into the system in time-slot $t$, and let $A(t_1, t_2)=\sum_{\tau=t_1}^{t_2}A(\tau)$ denote the cumulative packet arrivals to the system from time-slot $t_1$ to time-slot $t_2$. 

Next, we will introduce several assumptions on the arrival process for purpose of rate-function delay analysis.

\emph{\textbf{Assumption 1:}} The number of arrivals are bounded, i.e., there exists a finite number $L$ such that $A_i(t)\leq L$ for any $i$ and $t$. Also, we assume $\mathbb{P}(A(s, s+t-1)=Lnt)>0$ for any $s, t$ and $n$.

\emph{\textbf{Assumption 2:}} The arrival process are i.i.d across all users, and the mean arrival rate is $p$ (we assume $p<1$, otherwise the system could not be stable under any scheduling policy) for every user. Given any $\epsilon>0$ and $\delta>0$, there exists a positive function $I_B(\epsilon, \delta)$ independent of $n$ and $t$ such that
\begin{equation}
\mathbb{P}\Big{(}\frac{\Sigma_{\tau=1}^{t}\mathds{1}_{\{|A(\tau)-pn|>\epsilon n\}}}{t}>\delta\Big{)}<\exp(-ntI_B(\epsilon, \delta)).
\end{equation}
for all $t\geq T_B(\epsilon, \delta)$ and $n\geq N_B(\epsilon, \delta)$.

Assumptions 1 and 2 are mild. Packet arrivals per time-slot are typically bounded in practice. In addition, it has been shown in \cite{Lin_Tech_Report} that Assumption 2 is a general result of the statistical multiplexing effect of a large number of sources and holds for both \emph{i.i.d.} arrivals and Markov chain driven arrivals.
\begin{figure}[htbp]
  \centering
  \includegraphics[width=0.4\textwidth]{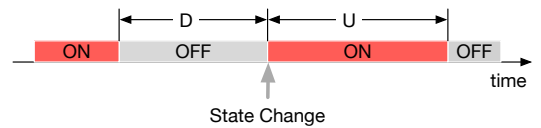}
  \caption{Time-correlated channel model}\label{fig:digit}
\end{figure}
\subsection{Stochastic Connectivity}
We assume that each channel has unit capacity and changes between ``ON" state and ``OFF" state from time to time. We use $C_{i, j}(t)$ to indicate the connectivity between queue $Q_i$ and server $S_j$ in time-slot $t$: $C_{i, j}(t)=1$ when the channel is ``ON" and $C_{i, j}(t)=0$ when the channel is ``OFF." We define ``ON" period to be the number of time-slots between the last time the channel was ``OFF" until the next time-slot it becomes ``OFF" again. ``OFF" period is defined in the similar way. From time to time, the channel state alternates between ``ON" periods and ``OFF" periods. We use an alternating renewal process to model the stochastic connectivity. In other words, 
the channel is initially ``ON" for a time period $U_1$ and then ``OFF" for a  time period $D_1$, followed by another ``ON" period $U_2$ and so on. In particular, the sequences of ``ON" times $\{U_n:n\geq 1\}$ and ``OFF" times $\{D_n:n\geq 1\}$ are independent sequences of \emph{i.i.d.} positive random variables. Let $U$ be a generic ``ON" time and $D$ be a generic ``OFF" time. We use $F_U(\cdot )$ and $F_D(\cdot)$ to denote the CDF of random variable $U$ and $D$, respectively. 

\emph{\textbf{Assumption 3:}} The sum of ``ON" and ``OFF" periods $U+D$ is aperiodic with $\mathbb{E}[U]<\infty$ and $\mathbb{E}[D]<\infty$.

It is well known (e.g. \cite{Renew_limit}) that under Assumption 3, we have:
\begin{align}
\lim_{t\to\infty}\mathbb{P}(C_{i, j}(t)=0)=\frac{\mathbb{E}[D]}{\mathbb{E}[U]+\mathbb{E}[D]}=\pi_0.\label{equ:Lim_C=0}
\end{align}
for any $i, j$.
\begin{rem} 
If $U+D$ is periodic with period $d$, the above result is true if $t$ is an integral multiple of $d$. For simplicity, we only focus on the aperiodic case.
\end{rem}

Note that this is a general model that can capture the time-correlation of a channel. If $U$ and $D$ are \textit{geometrically} distributed with parameters $1-q$ and $q$, respectively, it degenerates to a static \emph{i.i.d.} channel model with channel ``ON" probability $q$. Similarly, if $U$ and $D$ have a \textit{geometric} distribution with parameter $p_{10}$ and $p_{01}$, respectively,
it becomes the Markovian channel model with transition matrix $T=\left[
  \begin{array}{ccc}
    1-p_{01} & p_{01}\\
    p_{10} & 1-p_{10}\\
  \end{array}
\right]$.

In each time-slot, a scheduling policy allocates servers to serve packets from user queues. We further assume that a server can only serve one queue in a time-slot, however, a queue can get service from multiple servers simultaneously in one time-slot. In addition, one packet from queue $Q_i$ can be served if an ``ON" channel is allocated to queue $Q_i$.

\subsection{Problem Formulation}
In this paper, the metric we use to measure the delay performance is 
the large deviation rate-function of the steady-state probability that the largest packet delay exceeds a given threshold $b$. Assume the system starts at minus infinity, then $W(0)$ is the largest packet delay over all the queues in the steady-state. We define the rate-function $I(b)$ as the asymptotic decay-rate of the probability that $W(0)>b$ for a given threshold $b$:\begin{align}
I(b)\triangleq\lim_{n\to\infty}\frac{-1}{n}\log\mathbb{P}(W(0)>b).
\end{align}

Note that by the definition of rate-function $I(b)$, we can estimate the order of delay violation probability (i.e., $\mathbb{P}(W(0)>b)$ by $\exp(-nI(b))$. It is obvious that a larger rate-function implies a smaller delay violation probability and a better delay performance. In this paper, our objective is to maximize the rate-function $I(b)$. \footnote{We mainly focus on the delay analysis, since the results for throughput performance can be easily generalized from \cite{bo_hybrid}.}
\section{An Upper Bound On The rate-function}\label{Upper_bound}
In this section, we derive an upper bound on the best achievable delay rate-function. Later, we will use this upper bound as a baseline to evaluate the delay performance of the OPF policies. 

First, as in \cite{Lin_ITA, Lin_Tech_Report}, we define quantity $I_A(t, x)$ for any integer $t>0$ and any real number $x\geq 0$:
\begin{align}
I_A(t,x)\triangleq\sup_{\theta>0}[\theta(t+x)-\lambda_{A_i(-t+1, 0)}(\theta)].
\end{align}
where $\lambda_{A_i(-t+1, 0)}(\theta)=\log\mathbb{E}[e^{\theta A_i(-t+1, 0)}]$ is the cumulant-generating function of $A_i(-t+1,0)=\sum_{\tau=-t+1}^0A_i(\tau)$.

From Cramer's Theorem, $I_A(t,x)$ is equal to the asymptotic decay-rate of the probability that in any interval of $t$ time-slots, the total number of packet arrivals to the system is no smaller than $n(t+x)$ as $n$ tends to infinity, i.e., 
\begin{align}
\lim_{n\to\infty}\frac{-1}{n}\log\mathbb{P}(A(-t+1, 0)\geq n(t+x))=I_A(t, x).
\end{align}

We define $t_x$ for $L>1$ and non-negative integer $x$:
\begin{align}
t_x\triangleq\frac{x}{L-1}.
\end{align}

Then we define an integer set $\Psi_b\triangleq\{c\in\{1, 2, \cdots, b\}|t_{b-c}\in
\mathbb{Z}^{+}\}$. For any integer $b\geq 0$, let 
\begin{align}
I_U^{*}(b)&\triangleq\min\Big{\{}\log\big{(}\max_{\tau\in\{0, 1, \cdots\}}\frac{1-F_D(\tau)}{1-F_D(\tau+b)}\big{)}-\log\pi_0,\nonumber\\
&\min\big{\{}\inf_{t>t_{b}}I_A(t,b), \min_{1\leq c\leq b}\{\inf_{t>t_{b-c}}I_A(t, b-c)-\log\pi_0\nonumber\\
&+\log\big{(}\max_{\tau\in\{0, 1, \cdots\}}\frac{1-F_D(\tau)}{1-F_D(\tau+c-1)}\big{)}\big{\}},\nonumber\\
&\min_{c\in\Psi_b}\{I_A(t_{b-c}, b-c)-\log\pi_0+\nonumber\\
&\log\big{(}\max_{\tau\in\{0, 1, \cdots\}}\frac{1-F_D(\tau)}{1-F_D(\tau+c)}\big{)}\}\Big{\}}.
\end{align}

In addition, we define:
\begin{equation}
I_U(b)=
   \begin{cases}
   -\log\pi_0+\log\Big{(}\max_{\tau\in\{0, 1, \cdots\}}\frac{1-F_D(\tau)}{1-F_D(\tau+b)}\Big{)}, &\mbox{L=1}
   \\
   I_U^{*}(b), &\mbox{L\textgreater1}
   \end{cases}
\end{equation}

The following theorem shows that for any integer $b\geq 0$, $I_U(b)$ is an upper bound on the delay rate-function for any feasible scheduling policy.
\begin{thm}
For any integer threshold $b\geq 0$ and any scheduling policy, we have:
\begin{align}
\limsup_{n\to\infty}\frac{-1}{n}\log\mathbb{P}(W(0)>b)\leq I_U(b).
\end{align}
\end{thm}
\begin{IEEEproof}
We consider two cases $L>1$ and $L=1$. For the case $L>1$, we will consider three types of events: $\chi_1$, $\chi_2^{c}$ and $\chi_3^{c}$, which are subsets of the delay-violation event $\{W(0)>b\}$. Note that bursty arrivals and sluggish services both cause large packet delay in the system. In particular, $\chi_1$ is the event with sluggish services while $\chi_2^c$ and $\chi_3^c$ are events with bursty arrivals and sluggish services.
\ifreport

\emph{\textbf{Event $\chi_1$:}}
Suppose there is a packet that arrives to the network at the beginning of time-slot $-b-1$. Since the arrival process is independent across all queues, without loss of generality, we assume that this packet arrives to $Q_1$. Furthermore, $Q_1$ is assumed to be disconnected from all $n$ servers from time-slot $-b-1$ to $-1$. As a result, at the beginning of time-slot 0, this packet is still in the network and has a delay of $b+1$, which violates the delay threshold $b$. Therefore, $\chi_1\subseteq\{W(0)>b\}$. 

First we want to calculate the probability that $Q_1$ is disconnected from an arbitrary server $S_j$ from time-slot $-b-1$ to $-1$, i.e., $C_{1,j}(t)=0$ for all $-b-1\leq t\leq -1$. In this case, time-slot $-b-1$ must fall in some ``OFF" period and this ``OFF" period must cover the time from time-slot $-b-1$ to time-slot $-1$. Let $\hat{T}(-b, C_{i,j})$ be the length of the time period from the beginning of channel $C_{i,j}$'s last state-change ("ON" to "OFF") time-slot before time-slot $-b$ to the beginning of time-slot $-b-1$. Thus, we have:
\begin{align}
&\mathbb{P}(C_{1,j}(t)=0 \text{ for all }-b-1\leq t\leq -1)\nonumber\\
&=\sum_{\tau=0}^{\infty}\mathbb{P}(C_{1, j}(-b-1)=0, \hat{T}(-b, C_{1, j})=\tau, D\geq \tau+b+1)\nonumber\\
&=\sum_{\tau=0}^{\infty}\mathbb{P}(C_{1, j}(-b-1)=0)\cdot\mathbb{P}(\hat{T}(-b, C_{1, j})=\tau, \nonumber\\
&\qquad \qquad \qquad \qquad D\geq \tau+b+1|C_{1, j}(-b-1)=0)\nonumber\\
&\stackrel{(a)}=\pi_0\sum_{\tau=0}^{\infty}\mathbb{P}(\hat{T}(-b, C_{1, j})=\tau,\nonumber\\
&\qquad\qquad\qquad D\geq \tau+b+1|C_{1, j}(-b-1)=0)\nonumber\\
&=\pi_0\sum_{\tau=0}^{\infty}\mathbb{P}( D\geq \tau+b+1|\hat{T}(-b, C_{1, j})=\tau, C_{1, j}(-b-1)=0)\nonumber\\
&\qquad \qquad \qquad \cdot\mathbb{P}(\hat{T}(-b, C_{1, j})=\tau|C_{1, j}(-b-1)=0)\nonumber\\
&=\pi_0\sum_{\tau=0}^{\infty}\mathbb{P}( D\geq \tau+b+1|D\geq \tau+1)\cdot\nonumber\\
&\qquad \qquad \qquad \mathbb{P}(\hat{T}(-b, C_{1, j})=\tau|C_{1, j}(-b-1)=0)\nonumber\\
&\geq\pi_0\sum_{\tau=0}^{\infty}\frac{\mathbb{P}(D\geq \tau+b+1)}{\mathbb{P}(D\geq \tau+1)}\cdot\nonumber\\
&\qquad\qquad\qquad\mathbb{P}(\hat{T}(-b, C_{1, j})=\tau|C_{1, j}(-b-1)=0)\nonumber\\
&\geq\pi_0\min_{\tau\in\{0, 1, \cdots\}}\frac{1-F_D(\tau+b)}{1-F_D(\tau)}.
\end{align}
where (a) is from (\ref{equ:Lim_C=0}) since the system starts from $-\infty$ and is now in the steady-state.

Since all channels are independent from each other, we can easily obtain the probability for event $\chi_1$:
\begin{align}
\mathbb{P}(\chi_1)\geq \Big{(}\pi_0\cdot\min_{\tau\in\{0, 1, \cdots\}}\frac{1-F_D(\tau+b)}{1-F_D(\tau)}\Big{)}^n.
\end{align}

Hence, we have:
\begin{align}
\mathbb{P}(W(0)>b)\geq \mathbb{P}(\chi_1)\geq \Big{(}\pi_0\cdot\min_{\tau\in\{0, 1, \cdots\}}\frac{1-F_D(\tau+b)}{1-F_D(\tau)}\Big{)}^n.
\end{align}
and thus
\begin{align}
\limsup_{n\to\infty}&\frac{-1}{n}\log\mathbb{P}(W(0)>b)\nonumber\\
&\leq -\log\pi_0+\log\Big{(}\max_{\tau\in\{0, 1, \cdots\}}\frac{1-F_D(\tau)}{1-F_D(\tau+b)}\Big{)}.
\end{align}
\emph{\textbf{Event $\chi_2^{c}$:}}
Consider any fixed $c \in\{0, 1, 2, \cdots, b\}$ and any $t > t_{b-c}$. Recall that $t_{b-c}=\frac{b-c}{L-1}$. Then, for all $t > t_{b-c}$, we have $b-c < (L-1)t$, and thus $I_A(t, b-c) = I_A^{+}(t, b-c)$ from Lemma 11 (Right continuity of function $I_A(t,x)$) in \cite{bo_hybrid}. Hence, for any fixed $\epsilon > 0$, there exists a $\delta > 0$ such that $I_A(t, b - c + \delta) \leq I_A^{+}(t, b - c) + \epsilon = I_A(t, b - c) + \epsilon$. Suppose that from time-slot $-t - b$ to $-b - 1$, the total number of packet arrivals to the system is greater than
or equal to $nt + n(b - c + \delta)$, and let $p_{(b-c+\delta)}$ denote the probability that this event occurs. Then, from
Cramer's Theorem, we have $\lim_{n\to\infty}\frac{-1}{n}\log p_{(b-c+\delta)} = I_A(t, b - c + \delta) \leq I_A(t, b - c) + \epsilon$. Clearly, the total number of packets that are served in any time-slot is no greater than $n$. For any fixed $\delta$, we have $n\delta \geq 1$ for large enough $n$ (when $n \geq \frac{1}{\delta}$). Hence, if the above event occurs, at the end of time-slot $-c-1$, the system contains at least one packet that arrived before time-slot $-b$. 

Without loss of generality, we assume that this packet is in $Q_1$. Now, assume that $Q_1$ is disconnected from all $n$ servers from time-slot $-c$ to $-1$, i.e., $C_{1,j}(t)=0$ for all $1\leq j\leq n$, $-c\leq t\leq -1$.
Then, at the beginning of time-slot 0, there is still a packet that arrived before time-slot $-b$. Thus, we have $W(0) > b$ in this case. This implies $\chi_2^c\subseteq\{W(0) > b\}$. 

\emph{\textbf{Case 1: $c=0$}}

In this case, the probability that event $\chi_2^c$ occurs can be computed by:
\begin{align}
\mathbb{P}(\chi_2^0)=p_{(b+\delta)}.
\end{align}

And since $\chi_2^0\subseteq\{W(0)>b\}$, we have:
\begin{align}
\limsup_{n\to\infty}&\frac{-1}{n}\log\mathbb{P}(W(0)>b)\leq I_A(t, b)+\epsilon. \label{Case1}
\end{align}

\emph{\textbf{Case 2: $c\geq1$}} 

Applying the same method we used to bound the event $\chi_1$, we have:
\begin{align}
&\mathbb{P}(C_{1,j}(t)=0\text{ for all }1\leq j\leq n, -c\leq t\leq -1)\nonumber\\
&\geq\Big{(}\pi_0\cdot\min_{\tau\in\{0, 1, \cdots\}}\frac{1-F_D(\tau+c-1)}{1-F_D(\tau)}\Big{)}^n.
\end{align}

Note that the channel condition is independent from the arrival process, and the probability $\chi_2^c$ occurs can be bounded as:
\begin{align}
\mathbb{P}(\chi_2^c)\geq p_{(b-c+\delta)}\Big{(}\pi_0\cdot\min_{\tau\in\{0, 1, \cdots\}}\frac{1-F_D(\tau+c-1)}{1-F_D(\tau)}\Big{)}^n.
\end{align}

Hence, we have:
\begin{align}
\limsup_{n\to\infty}&\frac{-1}{n}\log\mathbb{P}(W(0)>b)\nonumber\\
&\leq I_A(t, b-c)+\epsilon-\log\pi_0\nonumber\\
&+\log\Big{(}\max_{\tau\in\{0, 1, \cdots\}}\frac{1-F_D(\tau)}{1-F_D(\tau+c-1)}\Big{)}. \label{Case2}
\end{align}

Since inequality (\ref{Case1}) or (\ref{Case2}) holds for any $c\in\{0, 1, 2, \cdots, b\}$, for any $t > t_{b-c}$, and any $\epsilon > 0$, applying the results we have in (\ref{Case1}) and (\ref{Case2}) and by letting $\epsilon$ tend to 0, taking the infimum over all $t > t_{b-c}$, and taking the minimum over all $c \in\{0, 1, 2, \cdots, b\}$, we have:
\begin{align}
&\limsup_{n\to\infty}\frac{-1}{n}\log\mathbb{P}(W(0)>b)\nonumber\\
&\leq \min\Big{\{}\inf_{t>t_{b}}I_A(t,b), \min_{c\in\{1, 2, \cdots, b\}}\{\inf_{t>t_{b-c}}I_A(t, b-c)\nonumber\\
&-\log\pi_0+\log\Big{(}\max_{\tau\in\{0, 1, \cdots\}}\frac{1-F_D(\tau)}{1-F_D\left(\tau+c-1\right)}\Big{)}\Big{\}}.
\end{align}

\emph{\textbf{Event $\chi_3^c$:}}
Consider any fixed $c\in\Psi_b\triangleq\{c\in\{1, 2, \cdots, b\}|t_{b-c}\in
\mathbb{Z}^{+}\}$. Suppose that from time-slot $-t_{b-c}-b$ to $-b-1$, the total number
of packet arrivals to the system is equal to $nt_{b-c}+n(b-c) = nLt_{b-c}$, and let $p^{\prime}_{(b-c)}$ denote the probability that this event occurs. Note that the total number of packet arrivals to the system from time-slot $-t_{b-c}-b$ to $-b-1$ can never exceed $nLt_{b-c}$. Then, from Cramer's Theorem, we have $\lim_{n\to\infty}\frac{-1}{n}\log p^{\prime}_{(b-c)}=I_A(t_{b-c}, b-c)$. Clearly, the total number of packets that can be served during the interval $[-t_{b-c}-b,-c-1]$
is no greater than $n(t_{b-c} + b - c) = nLt_{b-c}$. Suppose that there exists one queue, say $Q_1$ that is disconnected
from all the servers in time-slot $-c-1$. Then, at the end of time-slot
$-c - 1$, the system contains at least one packet that arrived before time-slot $-b$. Further, if queue $Q_1$ is disconnected from all the $n$ servers from time-slot $-c-1$ to $-1$. Then, at the beginning of time-slot $0$, we have a packet in the system that arrived before time-slot $-b$. Thus, in this case we have $W(0)>b$, and $\chi_3^c\subseteq\{W(0)>b\}$. Note that the probability that event $\chi_3^c$ occurs can be bounded as
\begin{align}
\mathbb{P}(\chi_3^c)\geq p^{\prime}_{(b-c)}\Big{(}\pi_0\cdot\min_{\tau\in\{0, 1, \cdots\}}\frac{1-F_D(\tau+c)}{1-F_D(\tau)}\Big{)}^n.
\end{align}

Since the above inequality holds for any $c\in\Psi_b$, by taking the minimum over all $c\in\Psi_b$, we have, for
\begin{align}
\limsup_{n\to\infty}&\frac{-1}{n}\log\mathbb{P}(W(0)>b)\nonumber\\
&\leq  \min_{c\in\Psi_b}\{I_A(t_{b-c}, b-c)-\log\pi_0\nonumber\\
&+\log\Big{(}\max_{\tau\in\{0, 1, \cdots\}}\frac{1-F_D(\tau)}{1-F_D(\tau+c)}\Big{)}\}.
\end{align}

Combining events $\chi_1$, $\chi_2^c$, $\chi_3^c$, the delay rate-function is upper bounded by $I_U^{*}(b)$:
\begin{align}
&\limsup_{n\to\infty}\frac{-1}{n}\log\mathbb{P}(W(0)>b)\nonumber\\
&\leq \min\Big{\{}\log\big{(}\max_{\tau\in\{0, 1, \cdots\}}\frac{1-F_D(\tau)}{1-F_D(\tau+b)}\big{)}-\log\pi_0,\nonumber\\
&\min\big{\{}\inf_{t>t_{b}}I_A(t,b), \min_{1\leq c\leq b}\{\inf_{t>t_{b-c}}I_A(t, b-c)-\log\pi_0\nonumber\\
&+\log\big{(}\max_{\tau\in\{0, 1, \cdots\}}\frac{1-F_D(\tau)}{1-F_D(\tau+c-1)}\big{)}\big{\}},\min_{c\in\Psi_b}\{I_A(t_{b-c}, b-c)\nonumber\\
&-\log\pi_0+\log\big{(}\max_{\tau\in\{0, 1, \cdots\}}\frac{1-F_D(\tau)}{1-F_D(\tau+c)}\big{)}\}\Big{\}}\nonumber\\
&\triangleq I_U^{*}(b).
\end{align}

Next, we consider the case where $L = 1$. In this case, we only need to consider event $\chi_1$, 
combining the case $L=1$ and $L>1$, we have shown that $I_U(b)$ is an upper bound on the delay rate-function.
\fi
\end{IEEEproof}
\section{Achievable Rate-function of OPF Policies}\label{Analysis}
In this section, we aim to derive a non-trivial achievable delay rate-function of the class of OPF policies. First, we state the definition of the class of OPF policies.
\begin{defn}
A scheduling policy \textbf{P} is said to be an OPF (\textit{oldest packets first}) policy if in any time-slot, policy \textbf{P} can serve the $k$ oldest packets in the system for the largest possible value of $k\in\{1, 2, \cdots, n\}.$
\end{defn}

We want to show that the achievable rate-function of any OPF policy $\textbf{P}$ is no smaller than $I_0(b)$, defined as:
\begin{align}
I_0(b)=
   \begin{cases}
   -\log\pi_0+b\cdot\log\frac{1}{1-\hat{q}}, &\mbox{L=1}
   \\
   I_0^{*}(b), &\mbox{L\textgreater1}
   \end{cases}
\end{align}
where the parameter $\hat{q}$ is defined to be 
\begin{align}
\hat{q}\triangleq\min\{\min_{k\in\{0, 1, \cdots\}}\frac{\mathbb{P}(D=k+1)}{1-F_D(k)}, \min_{k\in\{0, 1, \cdots\}}\frac{1-F_U(k+1)}{1-F_U(k)}\}.
\end{align}
and  \begin{align}
&I_0^{*}(b)\triangleq\min\Big{\{}b\cdot\log\frac{1}{1-\hat{q}}-\log\pi_0,\nonumber\\
&\min\{\inf_{t>t_b}I_A(t,b), \min_{1\leq c\leq b}\{\inf_{t>t_{b-c}}I_A(t, b-c)\nonumber\\
&\qquad-\log\pi_0+(c-1)\cdot\log\frac{1}{1-\hat{q}}\}\}\nonumber\\
&\min_{c\in\Psi_b}\{I_A(t_{b-c}, b-c)-\log\pi_0+c\cdot\log\frac{1}{1-\hat{q}}\}\Big{\}}.
\end{align}

The analysis of delay rate-function follows a similar line of argument as in the case of \emph{i.i.d.} channels. Specifically, we analyze the rate-function of the Frame Based Scheduling (FBS) policy and the perfect-matching policy and exploit the dominance property of the OPF policies over both of them. However, in the case of time-correlated channels, it becomes more challenging to derive a good lower bound on the achievable rate-function. Since the channel has different behaviors (distributions) for state-change and state-keeping. To address this key challenge, we prove two important properties of the FBS policy and the perfect-matching policy (Section IV.A), which will play a key role in the proof. We start by briefly describing the operations of the FBS policy and the perfect-matching policy.

Under the FBS policy, packets are served in unit of frames. Each frame is constructed according to a given operating parameter $h$, such that: 1) the difference of the arrival times of any two packets within a frame must be no greater than $h$; and 2) the total number of packets in each frame is no greater than $n_0=n-Lh$. In each time-slot, the packets arrived at the beginning of this time-slot are filled into the last frame until any of the above two conditions are violated, in which case a new frame will be opened. In each time-slot, the HOL frame can be served only if there exists a matching that can serve all the packets in the HOL frame. Otherwise, no packet will be served. In any time-slot, the FBS policy serves the HOL frame that contains the oldest (up to $n_0$) packets with high probability for a large $n$. 
Under the perfect-matching policy, if a perfect matching can be found, i.e., every queue can be matched with a different server that is connected to this queue, the HOL packet of every queue will be served by the respective server determined by the perfect matching.
Otherwise, none of the packets will be served. It has been shown in \cite{bo_hybrid} that any OPF policy dominates the FBS policy and the perfect-matching policy, i.e., given the same packet arrivals and channel realization, any OPF policy will serve every packet that the FBS policy has served up to time $t$; and the same for the perfect-matching policy. Therefore, the FBS and perfect-matching policy will provide lower bounds on the delay rate-function that any OPF policy can achieve.

\subsection{Properties of FBS and Perfect Matching Policy}\label{sec:FBS_PM}
In this subsection, we derive the following properties of FBS and perfect matching policy, which will later be used for the rate-function analysis. 
For ease of presentation,  we define function $X_F(t)$ as:
\begin{equation}
    X_F(t)=
   \begin{cases}
   1 &\mbox{if a frame can be served in time-slot $t$}\\ 
    &\mbox{\qquad under FBS policy,}\\
   0 &\mbox{otherwise.}
   \end{cases}
\end{equation}
We have the following lemma that gives a lower bound on the probability that $X_F(t)=1$.
\begin{lem}\label{Thm:perfect_matching}
Consider an $n\times n$ bipartite graph $G$, where the time-varying connectivity has the general time-correlation property described in Section \ref{Model}. Then, there exists an $N_{F}>0$, such that for all $n\geq N_{F}$ the conditional probability that $X_F(t)=1$ is bounded by:
\begin{align}
\mathbb{P}&(X_F(t)=1|\mathcal{S}(t_1), \cdots, \mathcal{S}(t_{d}), \mathcal{S}(t-1))\nonumber\\
&\geq1-\big{(}\frac{n}{1-\hat{q}}\big{)}^{7H}e^{-n\log\frac{1}{1-\hat{q}}}.
\end{align}
for any positive integer $d$, $t_1<t_2<\cdots<t_{d}<t-1$, and any $\mathcal{S}(t_1), \cdots, \mathcal{S}(t_{d})$ and all $n>N_F$, where $\mathcal{S}(\cdot)$ is the connectivity in the corresponding time-slot.
\end{lem}
\begin{IEEEproof}
We provide the proof in APPENDIX \ref{app_X_F}.
\end{IEEEproof}

Lemma \ref{Thm:perfect_matching} shows that given the past channel state information, a frame can be successfully served with high probability.
We are interested in finding an upper bound on the probability that during the time interval $[-t-b, -1]$, exactly $t+a$ frames can be successfully served by the FBS scheduling policy. We have the following lemma:
\begin{lem}\label{lem_sum}
For all $a\leq b-1$, we have:
\begin{align}
&\mathbb{P}(\sum_{\tau=-t-b}^{-1}X_F(\tau)=t+a)\nonumber\\
&\leq 2^{t+b}\big{(}\frac{n}{\pi_0}\big{)}^{7H}\big{(}\frac{n}{1-\hat{q}}\big{)}^{7bH}e^{-n\{-\log\pi_0+(b-a-1)\log\frac{1}{1-\hat{q}}\}}. \label{inequality_X_F}
\end{align}
\end{lem}
\begin{IEEEproof}
We provide the proof in APPENDIX \ref{app_sum}.
\end{IEEEproof}

Likewise, we define $X_{PM}$ as:
\begin{align}
   X_{PM}(t)=
   \begin{cases}
   1 &\mbox{if $G$ has a perfect matching at time-slot $t$,}\\
   0 &\mbox{otherwise.}
   \end{cases}
\end{align}
Similarly, we have the following lemma:
\begin{lem}\label{lemma_PM}
Consider an $n\times n$ bipartite graph $G$, where the time-varying connectivity has general time-correlation property. There exists an $N_{PM}>0$, for all $n\geq N_{PM}$ the probability that $G$ has no perfect matching can be bounded as:
\begin{align}
\mathbb{P}&(X_{PM}(t)=0|\mathcal{S}(t_1), \cdots, \mathcal{S}(t_{d}), \mathcal{S}(t-1))\nonumber\\
&\leq 3ne^{-n\log\frac{1}{1-\hat{q}}}.
\end{align}
\end{lem}
\begin{IEEEproof}
We omit the proof here, as the same technique used in the proof of Lemma \ref{Thm:perfect_matching} can be applied.
\end{IEEEproof}


Similarly, it can be shown that for all $a\leq b-1$:
\begin{align}
\mathbb{P}&(\sum_{\tau=-t-b}^{-1}X_{PM}(\tau)=t+a)\nonumber\\
&\leq 2^{t+3b}n^be^{-n\{-\log\pi_0+(b-a-1)\log\frac{1}{1-\hat{q}}\}}\label{inequality_X_PM}
\end{align}

The above inequality holds for sufficiently large $n\geq N_{PM}$. Note that the R. H. S. of inequalities (\ref{inequality_X_F}) and (\ref{inequality_X_PM}) are both monotonically increasing with respect to $a$.
\subsection{Achievable Rate-function}
We first consider the case where $L>1$. 
We need to pick an appropriate choice for the value of parameter $h$ for FBS based on the statistics of the arrival process. We fix $\delta<\frac{2}{3}$ and $\epsilon<p/2$. Then, from Assumption 2, there exists a positive function $I_B(\epsilon, \delta)$ such that for all $n\geq N_B(\epsilon, \delta)$ and $t\geq T_B(\epsilon, \delta)$, we have
\begin{equation}
\mathbb{P}\Big{(}\frac{\sum_{\tau=l+1}^{l+t}\mathds{1}_{\{|A(\tau)-pn|>\epsilon n\}}}{t}>\delta\Big{)}<\exp(-ntI_B(\epsilon, \delta)).
\end{equation}
where $l$ is any arbitrary integer. Choose parameter $h$ to be:
\begin{equation}
h=\max\Big{\{}T_B(\epsilon, \delta), \Big{\lceil}\frac{1}{(p-\epsilon)(1-\frac{3\delta}{2})}\Big{\rceil}, \Big{\lceil}\frac{2I_0(b)}{I_B(\epsilon, \delta)}\Big{\rceil}\Big{\}}+1.\label{equ:parameter_h}
\end{equation}
and define $H=Lh$.

The reason for choosing this value of $h$ will later become clearer. Note that in Assumption 2, the maximum number of arrivals in a time-slot is $L$.

Let $L(-b)$ be the last time before time-slot $-b$, when the backlog is empty, i.e., all the queues have a
queue-length of zero. Also, let $\mathcal{E}_t$ be the set of sample paths such that $L(-b) = -t-b-1$ and $W(0) > b$ under policy \textbf{P}. Then, we have
\begin{equation}
\mathbb{P}(W(0) > b)=\sum_{t=1}^\infty\mathbb{P}(\mathcal{E}_t).
\end{equation}

Let $\mathcal{E}_t^F$ and $\mathcal{E}_t^{PM}
$ be the set of sample paths such that given $L(-b) = -t-b-1$, the event $W(0) > b$ occurs under the FBS policy and the perfect-matching policy, respectively. Recall that policy \textbf{P} dominates both the FBS policy and the perfect-matching policy. Since each packet not served by the OPF policy is also not served by the FBS policy or perfect matching policy, then for any $t > 0$ we have
\begin{equation}\label{equ:subsetF_PM}
\mathcal{E}_t\subseteq\mathcal{E}_t^{F}\cap\mathcal{E}_t^{PM}.
\end{equation}

Recall that $p$ is the mean arrival rate to a queue. Now, we choose any fixed real number  $\hat{p}\in(p, 1)$,
and fix a finite time $t^*$ as
\begin{equation}
t^*\triangleq\max\Big{\{}T_1, \Big{\lceil}\frac{I_0(b)}{I_{BX}}\Big{\rceil}, \max\{t_{b-c}|c\in\Psi_b\}\Big{\}},
\end{equation}
\ifreport
where $T_1$ is defined as:
\begin{equation}
T_1\triangleq\max\Big{\{}T_B(\hat{p}-p, \frac{1-\hat{p}}{6(L+2)}), \Big{\lceil}\frac{6}{1-\hat{p}}\Big{\rceil}\Big{\}}.
\end{equation}
and
\begin{equation}
I_{BX}\triangleq\min\Big{\{}\frac{(1-\hat{p})\log\frac{1}{1-\hat{q}}}{9}, I_B(\hat{p}-p, \frac{1-\hat{p}}{6(L+2)})\Big{\}}.
\end{equation}
\fi

Hence, if we let
\begin{equation}
P_1\triangleq\sum_{t=1}^{t^*}\mathbb{P}(\mathcal{E}_t^F\cap\mathcal{E}_t^{PM}),
\end{equation}
and
\begin{equation}
P_2\triangleq\sum_{t=t^*}^{\infty}\mathbb{P}(\mathcal{E}_t^F\cap\mathcal{E}_t^{PM}).
\end{equation}

From the relation in (\ref{equ:subsetF_PM}), we can bound $\mathbb{P}(\mathcal{E}_t)$ as:
\begin{equation}
\mathbb{P}(\mathcal{E}_t)\leq P_1+P_2.
\end{equation}

Hence, we can divide the rate-function analysis into two parts. In part 1, we show that there exists a finite $N_1 > 0$ such that for all $n\geq N_1$, we have
\begin{equation}
P_1\leq C_1n^{7(b+1)H}e^{-nI_0(b)}.
\end{equation}

Then, in part 2, we show that there exists a finite $N_2>0$ such that for all $n\geq N_2$,
\begin{equation}
P_2\leq4e^{-nI_0(b)}.
\end{equation}

By combining part 1 and part 2, there exists a finite $N\triangleq\max\{N_1, N_2\}$, such that for all $n\geq N$,
\begin{equation}
\mathbb{P}(W(0)>b)\leq\Big{(}C_1n^{7(b+1)H}+4\Big{)}e^{-nI_0(b)}.
\end{equation}

If we take logarithm and limit as n goes to infinity, we obtain $\liminf_{n\to\infty}\frac{-1}{n}\log\mathbb{P}(W(0)>b)\geq I_0(b)$, which is the desired result.

\ifreport
\subsection{Part 1}
In this section, we want to show that there exists a finite value $N_1$, such that for all $n\geq N_1$, we have:
\begin{equation}
P_1\leq C_1n^{7(b+1)H}e^{-nI_0(b)}.
\end{equation}

First we consider the case when $t<t^{*}$, let $\mathcal{E}_{t}^{\alpha}$ denote the set of sample paths in which there are at least $n$ arrivals seen by every $h-1$ time-slots in the time interval $[-t-b, -b-1]$. Let $\mathcal{E}_{t}^{\beta}$ be the set of sample paths such that $\frac{A(-t-b, -b-1)}{n_0}-\sum_{\tau=-t-b}^{-1}X_F(\tau)>0$.  According to \cite{Lin_Tech_Report}, due to the choice of parameter $h$, we have
\begin{equation}
\mathcal{E}_t^F\subseteq(\mathcal{E}_t^\alpha)^c\cup\mathcal{E}_t^\beta
\end{equation}
and also there exists $N_3>0$ and $C_2>0$ such that for all $n\geq N_3$,
\begin{equation}
\mathbb{P}(\mathcal{E}_t^\alpha)>1-C_2te^{-nI_0(b)}.\label{equ:h_alpha_bound}
\end{equation}

For the probability of $\mathcal{E}_t^\beta$ for each $t$, we can derive an upper bound on the probability of a large burst of arrivals during an interval of $t$ time-slots \cite{bo_hybrid}.
\begin{align}
&\mathbb{P}(A(-t+1, 0)>n_0(t+x))\\\nonumber
&=\mathbb{P}(A(-t+1, 0)\geq(n-H)(t+x)+1)\\
&\leq e^{-nI_A(t,x)}e^{(H(t+x)-1)\theta^*}.
\end{align}
for all $x\in[0, (L-1)t]$ and $\theta^*\triangleq\max\{\theta_1, \theta_2, \cdots, \theta_{t^*}\}$ where $\theta_t\triangleq \text{argmax}_{\theta}[\theta(t+x)-\lambda_{A_i(-t+1, 0)}(\theta)]$.
Recall that $t_x=\frac{x}{L-1}$. 

We first consider any $t\in\{1, 2, \cdots, t^*\}\backslash\{t_{b-c^{'}}|c^{'}\in\Psi_b\}$.
Let $c_t\in\mathbb{Z}$ be the smallest integer such that $t_{b-c_t}<t$ (i.e., $t_{b-c_t}<t<t_{b-c_t+1}$ or $c_t=0$ if $t>t_b$). Then for $z\in\{c_t, c_t+1, \cdots, b\}$, we have $t_{b-z}<t$, thus $t + b - z < Lt$ and for all $z'<c_t$ we have $t_{b-z'}>t$ and thus $t+b-z'>Lt$. From the properties we have derived for FBS and perfect matching policy in section \ref{sec:FBS_PM}, we have for all $n\geq N_F$:
\begin{align}
&\mathbb{P}(\mathcal{E}_t^\beta)\nonumber\\
&=\mathbb{P}\Big{(}\frac{A(-t-b, -b-1)}{n_0}-X_F(-t-b, -1)>0\Big{)}\nonumber\\
&=\sum_{a=0}^{t+b-c_t}\mathbb{P}\Big{(}\sum_{\tau=-t-b}^{-1}X_F(\tau)=a\Big{)}\mathbb{P}(A(-t-b, -b-1)>an_0)\nonumber\\
&\leq(t+b+1)\max_{0\leq a\leq t+b-c_t}\{\mathbb{P}\Big{(}\sum_{\tau=-t-b}^{-1}X_F(\tau)=a\Big{)}\nonumber\\
&\times\mathbb{P}(A(-t-b, -b-1)>an_0)\}\nonumber\\
&\leq(t+b+1)\max\Big{\{}\max_{0\leq a\leq t-1}\big{\{}\mathbb{P}\big{(}\sum_{\tau=-t-b}^{-1}X_F(\tau)=a\big{)}\big{\}},\nonumber\\ 
&\mathbb{P}(A(-t-b, -b-1)>(t+b)n_0), \nonumber\\
&\max_{0\leq a\leq b-c_t}\big{\{}\mathbb{P}(\sum_{\tau=-t-b}^{-1}X_F(\tau)=t+a)\nonumber\\
&\qquad\times\mathbb{P}(A(-t-b, -b-1)>(t+a)n_0)\big{\}}\Big{\}}
\end{align}
Applying the results in section \ref{sec:FBS_PM} here, and we know that these values are monotonic increasing with respect to $a$.
\begin{align}
&\mathbb{P}(\mathcal{E}_t^\beta)\nonumber\\
&\leq(t+b+1)\max\Big{\{}2^{t+b}\big{(}\frac{n^{(b+1)}}{\pi_0(1-\hat{q})^{b}}\big{)}^{7H}e^{-n\{-\log\pi_0+b\log\frac{1}{1-\hat{q}}\}}, \nonumber\\
&\max_{a\in\{0, \cdots, b-c_t, c_t=0\}}\mathbb{P}(A(-t-b, -b-1)>(t+a)n_0), \nonumber\\
&\max_{a\in\{0, \cdots, b-c_t, c_t\ne 0\}}\{\mathbb{P}(\sum_{\tau=-t-b}^{-1}X_F(\tau)=t+a)\nonumber\\
&\qquad\times\mathbb{P}(A(-t-b, -b-1)>(t+a)n_0)\}\Big{\}}\nonumber\\
&\leq(t+b+1)2^{t+b}\big{(}\frac{n^{(b+1)}}{\pi_0(1-\hat{q})^{b}}\big{)}^{7H}e^{(H(t+b)-1)\theta^{*}}\nonumber\\
&\times\max\{e^{-n\{-\log\pi_0+b\log\frac{1}{1-\hat{q}}\}}, e^{-nI_A(t,b)} ,\nonumber\\
&\max_{a\in\{0, \cdots, b-c_t, c_t\ne 0\}}\{e^{-n\{I_A(t, a)-\log\pi_0+(b-a-1)\log\frac{1}{1-\hat{q}}\}}\}\}\nonumber\\
&\leq C_3n^{7(b+1)H} \exp\Big{\{}-n\min\{-\log\pi_0+b\log\frac{1}{1-\hat{q}}, \min_{c_t=0}I_A(t, b)\nonumber\\
&\min_{z\in\{c_t, c_{t+1},  \cdots, b\}}\{I_A(t, b-z)-\log\pi_0+(z-1)\log\frac{1}{1-\hat{q}}\}\}\Big{\}}
\end{align}
where $C_3=(t^*+b+1)2^{t^*+b}\left(\frac{1}{\pi_0(1-\hat{q})^{b}}\right)^{7H}e^{(H(t^*+b)-1)\theta^*}$, and we let $z=b-a$ in the last step. Hence, for all $n\geq N_4\max\{N_3, N_F\}$, we have
\begin{align}
\mathbb{P}(\mathcal{E}_t^F\cap\mathcal{E}_t^{PM})&\nonumber\leq\mathbb{P}(\mathcal{E}_t^F)\nonumber\\
&\leq1-\mathbb{P}(\mathcal{E}_t^\alpha)+\mathbb{P}(\mathcal{E}_t^\beta)\nonumber\\
&\leq C_4n^{7(b+1)H}e^{-nI_0(b)}
\end{align}
where $C_4\triangleq\max\{C_2t^*, C_3\}$ and $t\in\{1, 2, \cdots, t^*\}\backslash\{t_{b-c^{'}}|c^{'}\in\Psi_b\}$.

Next, we need to deal with any $t_{b-c}\in\{t_{b-c'}|c'\in\Psi_b\}$. In this case, we can use the dominance property over FBS and perfect matching policy, i.e., $\mathcal{E}_t\subseteq\mathcal{E}_t^F$ and $\mathcal{E}_t\subseteq\mathcal{E}_t^{PM}$. Note that we have $t_{b-c}=\frac{b-c}{L-1}>0$, thus
\begin{align}
t_{b-c}+b-c=Lt_{b-c}
\end{align}

If we define
\begin{align}
&\nonumber K_1\triangleq\mathbb{P}(\mathcal{E}_{t_{b-c}}^F\cap\mathcal{E}_{t_{b-c}}^{PM}, A(-t_{b-c}-b, -b-1)<(t_{b-c}+b-c)n_0)\\
&K_2\triangleq\mathbb{P}(\mathcal{E}_{t_{b-c}}^F\cap\mathcal{E}_{t_{b-c}}^{PM}, A(-t_{b-c}-b, -b-1)\geq(t_{b-c}+b-c)n_0)
\end{align}

According to union bound, we have:
\begin{equation}
\mathbb{P}(\mathcal{E}_{t_{b-c}}^F\cap\mathcal{E}_{t_{b-c}}^{PM})\leq K_1 + K_2\label{equ:union_bound}
\end{equation}

In particular,
\begin{align}
K_1&\nonumber\leq\mathbb{P}(\mathcal{E}_{t_{b-c}}^{F}, A(-t_{b-c}-b, -b-1)<(t_{b-c}+b-c)n_0)\\
&\leq1-\mathbb{P}(\mathcal{E}_{t_{b-c}}^\alpha)+K_1^{'},
\end{align}
where $K_1^{'}\triangleq\mathbb{P}(\mathcal{E}_{t_{b-c}}^{\beta}, A(-t_{b-c}-b, -b-1)<(t_{b-c}+b-c)n_0)$.

We have the similar result for $K_1^{'}$, $n\geq N_F$:
\begin{align}
K_1^{'}&=\mathbb{P}(\frac{A(-t_{b-c}-b, -b-1)}{n_0}-\sum_{\tau=-t_{b-c}-b}^{-1}X_F(\tau)>0, \nonumber\\
&\qquad\qquad A(-t_{b-c}-b, -b-1)<(t_{b-c}+b-c)n_0)\nonumber\\
&\leq\sum_{a=0}^{t_{b-c}+b-c-1}\Big{(}\mathbb{P}\big{(}\sum_{\tau=-t_{b-c}-b}^{-1}X_F(\tau)=a\big{)}\nonumber\\
&\qquad\qquad\times\mathbb{P}\big{(}A(-t_{b-c}-b, -b-1)>an_0\big{)}\Big{)}\nonumber\\
&\leq(t_{b-c}+b-c)\max_{0\leq a\leq t_{b-c}+b-c-1}\mathbb{P}\big{(}\sum_{\tau=-t_{b-c}-b}^{-1}X_F(\tau)=a\big{)}\nonumber\\
&\qquad\qquad\times\mathbb{P}\big{(}A(-t_{b-c}-b, -b-1)>an_0\big{)}\nonumber\\
&\leq C_3n^{7(b+1)H}\exp\Big{\{}-n\min\{-\log\pi_0+b\log\frac{1}{1-\hat{q}}, \nonumber\\
&\qquad\min_{c\leq z\leq b}\{I_A(t, b-z)-\log\pi_0+(z-1)\log\frac{1}{1-\hat{q}}\}\}\Big{\}}
\end{align}

For $K_2$, we have
\begin{align}
K_2&\nonumber\triangleq\mathbb{P}(\mathcal{E}_{t_{b-c}}^F\cap\mathcal{E}_{t_{b-c}}^{PM}, A(-t_{b-c}-b, -b-1)\geq(t_{b-c}+b-c)n_0)\\
&\leq\mathbb{P}(\mathcal{E}_{t_{b-c}}^{PM}, A(-t_{b-c}-b, -b-1)\geq(t_{b-c}+b-c)n_0)
\end{align}

For all $n\geq N_{PM}$
\begin{align}
K_2&\nonumber\leq\mathbb{P}(\mathcal{E}_{t_{b-c}}^{PM}|A(-t_{b-c}-b, -b-1)\geq(t_{b-c}+b-c)n_0)\nonumber\\
&\qquad\qquad\times\mathbb{P}(A(-t_{b-c}-b, -b-1)\geq(t_{b-c}+b-c)n_0)\nonumber\\
&\leq\mathbb{P}(\mathcal{E}_{t_{b-c}}^{PM}|A(-t_{b-c}-b, -b-1)=Ln_0t_{b-c})\nonumber\\
&\qquad\qquad\times e^{-nI_A(t_{b-c}, b-c)}e^{(H(t_{b-c}+b-c)-1)\theta^*}\nonumber\\
&\leq\mathbb{P}(\sum_{\tau=-t_{b-c}-b}^{-1}X_{PM}(\tau)<t_{b-c}+b-c)\nonumber\\
&\qquad\qquad\times e^{-nI_A(t_{b-c}, b-c)}e^{(H(t_{b-c}+b-c)-1)\theta^*}\nonumber\\
&\leq \sum_{a=0}^{t_{b-c}+b-c-1}\mathbb{P}(\sum_{\tau=-t_{b-c}-b}^{-1}X_{PM}(\tau)=a)\nonumber\\
&\qquad\qquad\times e^{-nI_A(t_{b-c}, b-c)}e^{(H(t_{b-c}+b-c)-1)\theta^*}\nonumber\\
&\leq(t_{b-c}+b-c)\nonumber\\
&\qquad\times\max_{a\in\{0, \cdots, t_{b-c}+b-c-1\}}\mathbb{P}(\sum_{\tau=-t_{b-c}-b}^{-1}X_{PM}(\tau)=a)\nonumber\\
&\qquad\times e^{-nI_A(t_{b-c}, b-c)}e^{(H(t_{b-c}+b-c)-1)\theta^*}\nonumber\\
&\leq(t_{b-c}+b-c)2^{t+3b}e^{(H(t_{b-c}+b-c)-1)\theta^*}n^b\nonumber\\
&\qquad\times e^{-n(I_A(t_{b-c}, b-c)-\log\pi_0+c\log\frac{1}{1-\hat{q}})}\nonumber\\
&\leq C_5 n^be^{-n(I_A(t_{b-c}, b-c)-\log\pi_0+c\log\frac{1}{1-\hat{q}})}
\end{align}
where $C_5\triangleq(t^*+b)2^{t^*+3b}e^{(H(t^*+b)-1)\theta^*}$.

Combining results in (\ref{equ:h_alpha_bound}) and (\ref{equ:union_bound}), we have for any $t_{b-c}\in\{t_{b-c}|c\in\Psi_b\}$ and $n\geq N_5\triangleq\max\{N_3, N_F, N_{PM}\}$,
\begin{align}
\mathbb{P}(\mathcal{E}_{t_{b-c}}^F\cap\mathcal{E}_{t_{b-c}}^{PM})\leq C_6n^{7(b+1)H}e^{-nI_0(b)},
\end{align}
where $C_6\triangleq\max\{C_2t^*, C_3, C_5\}$.

Summing over $t=1$ to $t=t^*$, we have
\begin{align}
P_1&=\sum_{t=1}^{t^*}\mathbb{P}(L(-b)=-t_{b-c}-b-1, \mathcal{E}_t)\\
&\leq C_1n^{7(b+1)H}e^{-nI_0(b)},
\end{align}
for all $n\geq N_1\triangleq\max\{N_4, N_5\}$, where $C_1\triangleq C_6 t^*$.
\subsection{Part 2}
In this section, we want to show that there exists an $N_2>0$, such that for $n\geq N_2$,
\begin{equation}
P_2\leq 4e^{-nI_0(b)}
\end{equation}

Let $R_0$ be the empty space in the end-of-line frame at the end of time-slot $t_1$. Then, let $A^{R_0}_F (t_1, t_2)$
denote the number of new frames created from time-slot $t_1$ to $t_2$, including any partially-filled frame in
time-slot $t_2$, but excluding the partially-filled frame in time-slot $t_1$. Also, let $A_F (t_1, t_2) = A^{R_0}_F (t_1, t_2)$, if
$R_0 = 0$. As in the proof for Theorem 2 of \cite{Lin_Tech_Report}, for any fixed real number $\hat{p}\in(0,1)$, we consider the
arrival process $\hat{A}(\cdot)$, by adding extra dummy arrivals to the original arrival process ${A}(\cdot)$. The resulting
arrival process $\hat{A}(\cdot)$ is simple, and has the following property:
\begin{equation}
    \hat{A}(\tau)=
   \begin{cases}
   \hat{p}n &\mbox{if $A(\tau)\leq\hat{p}n$}\\
   Ln &\mbox{if $A(\tau)>\hat{p}n$}
   \end{cases}
\end{equation}

Since $\hat{A}(\tau)\geq{A}(\tau)$, if we can find an upper bound on $\hat{A}_F(-t-b, -b-1)$, then it is also an upper bound on $A_F(-t-b, -b-1)$.

Consider any $t\geq t^{*}$. Let $B=\{b_1, b_2, \cdots, b_{|B|}\}$ be the set of time-slots in the interval from $-t-b$ to $-b-1$ when $\hat{A}(\tau)=Ln$. Given $L(-b)=-t-b-1$, from Corollary 2 of \cite{Lin_ITA} and the proof of Theorem 2 in \cite{bo_hybrid}, we have
\begin{equation}
\hat{A}_F(-t-b, -b-1)\leq\frac{n}{n_0}(\hat{p}t+(L+2)|B|+1)
\end{equation}

From Assumption 2, we know that $|B|$ could be arbitrary small for large enough $t$ and $n$. Suppose for $n\geq\frac{(2+\hat{p})H}{1-\hat{p}}$, $t>\frac{6}{1-p}$ and $|B|<\frac{1-\hat{p}}{6(L+2)}t$
\begin{align}
A_F(-t-b, -b-1)&\nonumber\leq\hat{A}_F(-t-b, -b-1)\\\nonumber
&\leq\frac{n}{n_0}(\hat{p}t+(L+2)|B|+1)\\\nonumber
&<\frac{2+\hat{p}}{1+2\hat{p}}\Big{(}\hat{p}t+\frac{1-\hat{p}}{3}t\Big{)}\\
&\leq(\frac{2+\hat{p}}{3})t.
\end{align}

In fact, for $n\geq N_6\triangleq\max\{N_B(\hat{p}-p, \frac{1-\hat{p}}{6(L+2)}), \frac{(2+\hat{p})H}{1-\hat{p}}\}$ and $t\geq T_1$.
\begin{align}
&\mathbb{P}(A_F(-t-b, -b-1)\nonumber\\
&\geq(\frac{2+\hat{p}}{3})t, L(-b)=-t-b-1)\nonumber\\
&=1-\mathbb{P}(A_F(-t-b, -b-1)<(\frac{2+\hat{p}}{3})t, \nonumber\\
&\qquad L(-b)=-t-b-1)\nonumber\\
&\leq1-\mathbb{P}(|B|\leq\frac{1-\hat{p}}{6(L+2)}t)\\
&\leq e^{-ntI_B(\hat{p}-p, \frac{1-\hat{p}}{6(L+2)})},\label{equ:A_Fbig}
\end{align}
where the last inequality is from Assumption 2.

Next, let's restate Lemma 1 in \cite{Lin_Tech_Report},
\begin{lem}\label{lem:chernoff_bound}
Let $X_i$, $i=1, 2, \cdots$ be a sequence of binary random variables such that for all $i$,
\begin{equation}
\mathbb{P}\Big{(}X_i=0|X_{i'}, i'< i\Big{)}\leq c(n)e^{-nb}\label{new_cond}
\end{equation}
where $c(n)$ is a polynomial in $n$ of finite degree. Let $N_9$ be
such that $c(n) < e^{\frac{nb}{2}}$ for all $n > N_9$. Then, for any $0 < a < 1$,
\begin{equation}
\mathbb{P}\Big{(}\sum_{i=1}^tX_i<(1-a)t\Big{)}<e^{-\frac{tnab}{3}}
\end{equation}
for all $n>N_{10}:=\max\{\frac{12}{ab}, N_9\}$.
\end{lem}
\begin{IEEEproof}
Due to the choice of $N_9$, we have for all $n>N_9$,
\begin{align}
\mathbb{P}\Big{(}X_i=0|X_{i'}, i'< i\Big{)}\leq e^{-\frac{nb}{2}}.
\end{align}

From Markov's inequality we have for any $r>0$:
\begin{align}
\mathbb{P}\Big{(}\sum_{i=1}^tX_i\leq(1-a)t\Big{)}&=\mathbb{P}\Big{(}e^{-r\sum_{i=1}^tX_i}\geq e^{-r(1-a)t}\Big{)}\nonumber\\
&\leq\frac{\mathbb{E}[e^{-r\sum_{i=1}^tX_i}]}{e^{-r(1-a)t}}
\end{align}

From the definition of conditional expectation, we have:
\begin{align}
\mathbb{E}[e^{-r\sum_{i=1}^tX_i}]&=\mathbb{E}[\mathbb{E}[e^{-r\sum_{i=1}^tX_i}|X_1, X_2, \cdots, X_{t-1}]]\nonumber\\
&=\mathbb{E}[e^{-r\sum_{i=1}^{t-1}X_i}\mathbb{E}[e^{-rX_t}|X_1, X_2, \cdots, X_{t-1}]]\nonumber\\
&\leq\mathbb{E}[e^{-r\sum_{i=1}^{t-1}X_i}]\Big{(}(1-e^{-\frac{nb}{2}}) + e^{-r-\frac{nb}{2}}\Big{)}
\end{align}

The last inequality comes from the condition (\ref{new_cond}), continuing for $t-1, t-2, \cdots$ down to 1, we have:
\begin{align}
\mathbb{E}[e^{-r\sum_{i=1}^tX_i}]\leq\Big{(}(1-e^{-\frac{nb}{2}}) + e^{-r-\frac{nb}{2}}\Big{)}^{t}
\end{align}

Applying the proof technique used in \cite{{IEEEproof_Chernoff}}, we can get:
\begin{align}
\mathbb{P}\Big{(}\sum_{i=1}^tX_i\leq(1-a)t\Big{)}<e^{-tD(a||e^{-\frac{nb}{2}})}
\end{align}

From the proof of Lemma 1 in \cite{Lin_Tech_Report}, we know that the result holds for sufficiently large $n>N_{10}$.
\end{IEEEproof}

From Lemma \ref{lem:chernoff_bound}, there exists $N_7\triangleq\max\{N_{10}, N_F\}$, such that for all $n\geq N_7$,
\begin{align}
&\nonumber\mathbb{P}(X_F(-t-b, -b-1)<(\frac{2+\hat{p}}{3})t, \nonumber\\
&\qquad\qquad L(-b)=-t-b-1)\nonumber\\
&\leq\mathbb{P}(X_F(-t-b, -b-1)<(\frac{2+\hat{p}}{3})(t+b), \nonumber\\
&\qquad\qquad L(-b)=-t-b-1)\nonumber\\
&\leq e^{-n(t+b)(\frac{1-\hat{p}}{9}\log\frac{1}{1-\hat{q}}})\nonumber\\
&\leq e^{-nt\frac{(1-\hat{p})\log\frac{1}{1-\hat{q}}}{9}}\label{equ:X_Fsmall}
\end{align}

From (\ref{equ:A_Fbig}) (\ref{equ:X_Fsmall}), for all  $n\geq N_8\triangleq\max\{N_6, N_7\}$ and $t\geq T_1$
\begin{align}
&\nonumber\mathbb{P}\Big{(}A_F(-t-b, -b-1)-X_F(-t-b ,-1)>0, \nonumber\\
&\qquad\qquad L(-b)=-t-b-1\Big{)}\nonumber\\
&\leq 1-(1-e^{-ntI_B(\hat{p}-p, \frac{1-\hat{p}}{6(L+2)})})(1-e^{-nt\frac{(1-\hat{p})\log\frac{1}{1-\hat{q}}}{9}})\nonumber\\
&\leq2e^{-ntI_{BX}}
\end{align}

Summing over all $t\geq t^{*}$, we have for $n\geq N_2\triangleq\max\{N_8, \big{\lceil}\frac{\log 2}{I_{BX}}\big{\rceil}\}$,
\begin{align}
P_2&\nonumber=\sum_{t=t^{*}}^\infty\mathbb(\mathcal{E}_t^F\cap\mathcal{E}_t^{PM})\nonumber\\
&\leq\sum_{t=t^*}^\infty\mathbb{P}(\mathcal{E}_t^F)\nonumber\\
&\leq\sum_{t=t^*}^\infty\mathbb{P}\Big{(}A_F(-t-b, -b-1)-X_F(-t-b ,-1)>0, \nonumber\\
&\qquad\qquad L(-b)=-t-b-1\Big{)}\nonumber\\
&\leq\sum_{t=t^*}^\infty2e^{-ntI_{BX}}\nonumber\\
&\leq4e^{-nt^*I_{BX}}\nonumber\\
&\leq4e^{-nI_0(b)}
\end{align}
where the last two inequalities are from the choice of $N_2$ and $I_{BX}$.

Combining part 1 and part 2, it is easy to verify $\liminf_{n\to\infty}\frac{-1}{n}\log\mathbb{P}(W(0)>b)\geq I_0(b)$.

\subsection{Delay Rate-Function Analysis for $L=1$}
In this case, we want to show that for any fixed integer $b>0$, the rate-function achieved by OPF policy is greater than $b\log\frac{1}{1-\hat{q}}-\log\pi_0$.

Similarly, we can choose $t'$ as:
\begin{equation}
t'\triangleq\max\Big{\{}T_1, \Big{\lceil}\frac{b\log\frac{1}{1-\hat{q}}-\log\pi_0}{I_{BX}}\Big{\rceil}\Big{\}}
\end{equation}

If we define probabilities
\begin{align}
&P_1^{'}\triangleq\sum_{t=1}^{t'}\mathbb{P}(\mathcal{E}_t^F\cap\mathcal{E}_t^{PM})\\
&P_2^{'}\triangleq\sum_{t=t'}^{\infty}\mathbb{P}(\mathcal{E}_t^F\cap\mathcal{E}_t^{PM})
\end{align}

Due to the dominance property over FBS policy and the perfect-matching policy, we can spilt the delay violation probability $\mathbb{P}(W(0)>b)$ as:
\begin{equation}
\mathbb{P}(W(0)>b)\leq P_1^{'}+P_2^{'}.
\end{equation}

Likewise, we still divide the proof into two parts. In part 1, we need to show that for all $n\geq N_1^{'}$
\begin{equation}
P_1^{'}\leq C_1^{'}n^{b}e^{-n\big{(}b\frac{1}{1-\hat{q}}-\log\pi_0\big{)}}
\end{equation}

Consider $t\leq t'$, the total packet arrivals during the interval of $[-t-b, -b-1]$ cannot exceed $nt$ when $L=1$, hence if $\sum_{\tau=-t-b}^{-1}X_{PM}(\tau)\geq t$ then all packets that arrive before $-b$ have been served at time-slot $0$, event $\mathcal{E}_t^{PM}$ does not occur. We have $\mathcal{E}_t^{PM}\subseteq\big{\{}L(-b)=-t-b-1, \sum_{\tau=-t-b}^{-1}X_{PM}(\tau)<t\big{\}}$

Thus, we have
\begin{align}
&\nonumber\mathbb{P}(\mathcal{E}_t^F\cap\mathcal{E}_t^{PM})\\\nonumber
&\leq\mathbb{P}(\mathcal{E}_t^{PM})\\\nonumber
&\leq\mathbb{P}\Big{(}\sum_{\tau=-t-b}^{-1}X_{PM}(\tau)<t\Big{)}\\\nonumber
&\leq t\max_{a\in\{0,\cdots, t-1\}}\mathbb{P}\Big{(}\sum_{\tau=-t-b}^{-1}X_{PM}(\tau)=a\Big{)}\\\nonumber
&\leq t'2^{t'+3b}n^{b}e^{-n\big{(}b\log\frac{1}{1-\hat{q}}-\log\pi_0\big{)}}
\end{align}
where $C_3^{'}\triangleq t^{'}2^{t^{'}+3b}$.

Let $C_1^{'}\triangleq t'C_3^{'}$, summing over all $t\leq t'$, we have
\begin{align}
P_1^{'}\leq C_1^{'}n^be^{-n\big{(}b\log\frac{1}{1-\hat{q}}-\log\pi_0\big{)}}
\end{align}

For part 2, by applying the same argument as in the case of $L > 1$, we can show that there exists a finite $N_2^{'}$ such that for all $n\geq N_2^{'}$, we have $P_2^{'}\leq4e^{-n\big{(}b\log\frac{1}{1-\hat{q}}-\log\pi_0\big{)}}$.
Combining both parts, we complete the proof for the case of $L = 1$. 

Finally, combining both cases of $L > 1$ and $L = 1$, we show that the OPF policy achieves delay rate-function at least $I_0(b)$ in general time-correlated channel.
\fi
\section{The Relationship Between $I_U(b)$ and $I_0(b)$}\label{Relationship}
We have already shown that $I_U(b)$ is an upper bound on the delay rate-function under any possible scheduling policies. Also, we show that the delay rate-function that can be achieved by any OPF policy is no smaller than $I_0(b)$. 
In this section, we investigate the relationship between the values of these two rate-functions. We show that if the channel is \textit{non-negatively correlated} (Condition A) and the distribution of the OFF period is memoryless (Condition B), any OPF policy can achieve the optimal delay rate-function, i.e., $I_U(b)=I_0(b)$ for \emph{any} fixed integer $b \ge 0$. 
\subsubsection{Condition A: Any vector of finite channel states satisfies non-negative correlation condition}~\\
In statistics, two random variables $X,Y$ are \textit{non-negatively correlated} if $cov(X,Y)=\mathbb{E}[XY]-\mathbb{E}[X]\mathbb{E}[Y]\geq 0$. The following definition from \cite{Negative_regression} is a reasonable generalization of non-negative correlation to a set of random variables.
\begin{defn}
(\textbf{Non-negative Correlation Condition}) Let $\bold{X}=(X_1, \cdots, X_n)$ be a vector of random variables. Then the random vector $\bold{X}$ satisfies non-negative correlation condition if the conditional expectation $\mathbb{E}[X_i, i\in \mathbb{I}|X_{j}=t_{j}, \text{for }\forall{j}\in \mathbb{J}]$ is non-decreasing in each $t_{j}$, ${j}\in \mathbb{J}$ for any disjoint index set $\mathbb{I}, \mathbb{J}\subseteq[n]$. 
\end{defn}
\begin{lem}\label{tilde_q}
If condition A holds, then the class of OPF policies can achieve a delay rate-function of $I_0(b)$ with parameter $\hat{q}$ replaced by $\tilde{q}$, which is given by:
\begin{align} 
\tilde{q}=\min_{k\in\{0, 1, \cdots\}}\frac{\mathbb{P}(D=k+1)}{1-F_D(k)}.
\end{align}
\end{lem}
\begin{IEEEproof}
\ifreport
Note that condition A implies:
\begin{align}
\mathbb{E}[C_{i,j}(t)|C_{i,j}(t_1)=c_1, \cdots, C_{i,j}(t_p)=c_p, C_{i,j}(t-1)=1]\nonumber\\
\geq \mathbb{E}[C_{i,j}(t)|C_{i,j}(t_1)=c_1, \cdots, C_{i,j}(t_p)=c_p, C_{i,j}(t-1)=0]
\end{align}
for any $c_1, \cdots, c_p\in \{0, 1\}$.

Since $C_{i,j}(t)$ is a binary random variable, we can rewrite the inequality as follows:
\begin{align}\label{positive_prob}
\mathbb{P}(C_{i,j}(t)=1|\mathcal{S}(t_1)=\bold{s_1}, \cdots, \mathcal{S}(t_p)=\bold{s_p}, C_{i,j}(t-1)=1)\nonumber\\
\geq \mathbb{P}(C_{i,j}(t)=1|\mathcal{S}(t_1)=\bold{s_1}, \cdots, \mathcal{S}(t_p)=\bold{s_p}, C_{i,j}(t-1)=0)
\end{align}
where $\bold{s_1}, \cdots, \bold{s_p}$ are any arbitrary binary $n\times n$ matrixes. \\
Substituting inequality (\ref{positive_prob}) into (\ref{lemma_first}), the result follows.
\fi
\end{IEEEproof}

\subsubsection{Condition B: Distribution $D$ has a memoryless property}~
When the distribution $D$ has a memoryless property, namely $D$ is \textit{geometrically} distributed, we have:
\begin{align}
\frac{1-F_D(k+n-1)}{1-F_D(k+n)}&=\frac{1-F_D(k+n-2)}{1-F_D(k+n-1)}\nonumber\\
&=\cdots=\frac{1-F_D(k)}{1-F_D(k+1)}.
\end{align}
for any $k\geq 1$ and $1\leq n\leq b$. Multiplying all these $n$ fractions, we can obtain the following equation:
\begin{align}
\Big{(}\frac{1-F_D(k)}{1-F_D(k+1)}\Big{)}^n=\frac{1-F_D(k)}{1-F_D(k+n)}.
\end{align}

Finally, if the above two conditions are both satisfied, we have the following theorem:
\begin{thm}
The class of OPF policies achieve optimal delay rate-function performance under the general time-correlated channel model if conditions A and B both hold:
\end{thm}
\begin{IEEEproof}
Condition A ensures that $\tilde{q}$ is related to the distribution of random variable $D$ and does not depend on $U$. If we substitute the value of $\tilde{q}$ into $I_0(b)$, it is easy to see that the expression for $I_0(b)$ is very similar to $I_U(b)$, except for the terms related to the CDF of $D$. Applying condition B, we can obtain $I_0(b)\geq I_U(b)$ directly. Since $I_0(b)$ is an lower bound on the delay rate-function that can be achieved by any OPF policy and $I_U(b)$ is an upper bound on the delay rate-function under any possible scheduling policies, we can conclude that the class of OPF policies achieve delay rate-function optimality in general correlated channel model.
\end{IEEEproof}

In fact, \emph{i.i.d.} channel and \textit{non-negatively correlated} Markovian channel are two special cases, in which both conditions A and B are satisfied, and thus, the optimal rate-function is achieved.
\begin{rem}
Under \emph{i.i.d.} channel model with channel ``ON" probability $q$, conditions A and B always hold. In this case, $U$ has a \textit{geometric} distribution with parameter $1-q$, and $D$ has a \textit{geometric} distribution with parameter $q$. 
\begin{align}
\mathbb{E}[C_{i,j}(t)|C_{i,j}(t_1)=c_1, \cdots, C_{i,j}(t_k)=c_k]=\mathbb{E}[C_{i,j}(t)]=q. \nonumber
\end{align}
Since the conditional expectations remain the same for any $c$, the non-negative correlation condition (condition A) holds.
On the other hand, since random variable $D$ is \textit{geometrically} distributed, $D$ has a memoryless property, i.e., condition B holds.
\end{rem}

\begin{rem}
Under Markovian channel model with transition matrix $T$, condition A is equivalent to the standard notion of non-negative correlation for a two-state Markov chain. In this case, $U$ has a \textit{geometric} distribution with parameter $p_{10}$, and $D$ has a \textit{geometric} distribution with parameter $p_{01}$. Substituting the PMF of the \textit{geometric} distribution, we have
\begin{align}
&\mathbb{E}[C_{i,j}(t)|C_{i,j}(t_1)=c_1, \cdots, C_{i,j}(t-1)=1]\nonumber\\
&=\mathbb{P}(C_{i,j}(t)=1|C_{i,j}(t-1)=1)=1-p_{10}.
\end{align}
and 
\begin{align}
&\mathbb{E}[C_{i,j}(t)|C_{i,j}(t_1)=c_1, \cdots, C_{i,j}(t-1)=0]\nonumber\\
&=\mathbb{P}(C_{i,j}(t)=1|C_{i,j}(t-1)=0)=p_{01}.
\end{align}

Hence, condition A is equivalent to:
\begin{align}
1-p_{10}\geq p_{01}\iff p_{01}+p_{10}\leq 1.
\end{align} 
which is the condition for non-negative correlation in a two-state Markov chain. Similarly, condition B is satisfied because $D$ is also \textit{geometrically} distributed.
\end{rem}

\begin{thm}
Under \textit{negatively correlated} Markovian channel model, i.e., $p_{01}+p_{10}>1$, the class of OPF policies can achieve a delay rate-function that is no smaller than $\frac{\log p_{10}}{\log(1-p_{01})}$-fraction of the optimal value, where $p_{01}$ and $p_{10}$ come from the transition probability.
\end{thm}
\begin{IEEEproof}
Since $p_{01}+p_{10}>1$, the conditional probability $\mathbb{P}(C_{i,j}(t)=1|C_{i,j}(t-1))$ is lower bounded by $1-p_{10}$. Thus, by using the same proof technique, we can show that the same results hold for $\log\frac{1}{1-p_{01}}$ replaced by $\log\frac{1}{p_{10}}$. Note that the upper bound still remains the same, therefore, it is easy to see that the delay rate-function achieved by the OPF policies is no smaller than $\frac{\log p_{10}}{\log(1-p_{01})}$-fraction of the optimal value.
\end{IEEEproof}

\section{Numerical Results}\label{Simulation}
In this section, we conduct simulations to compare scheduling performance under different channel settings. Among all the OPF policies such as delay weighted matching (DWM) \cite{Lin_ITA, Lin_Tech_Report}, DWM-n and hybrid policy\cite{bo_hybrid}, we choose DWM in
our simulations as DWM has the best empirical performance
in various scenarios \cite{Lin_Tech_Report}. The DWM policy considers at most $n$ oldest packets from each queue, i.e., a total of at most $n^2$ packets and chooses the schedule that maximizes the sum of
the delays in each time-slot. 
We consider 0-5 \emph{i.i.d.} arrivals i.e., \begin{equation}
A_i(t)=
   \begin{cases}
   5, &\mbox{with probability $\mu$,}\\ 
   0, &\mbox{with probability $1-\mu$,}
   \end{cases}
\end{equation}
for all $i$. The arrival processes are assumed to be independent
across all the queues. For the channel model, we
assume that all the channels are homogeneous and consider
the following seven channel settings, channel settings
1 and 2 are \emph{i.i.d.} ON/OFF channels with ``ON'' probability $q_1=0.6$ and $q_2=0.5$, respectively, and channel settings
3, 4, 5, 6 and 7 are Markovian channels with transition matrix $T=\left[
\begin{array}{ccc}
    0.94 & 0.06\\
    0.04 & 0.96\\
  \end{array}
\right]$, $\left[
\begin{array}{ccc}
    0.85 & 0.15\\
    0.1 & 0.9\\
  \end{array}
\right]$, $\left[
\begin{array}{ccc}
    0.01 & 0.99\\
    0.99 & 0.01\\
  \end{array}
\right]$, $\left[
\begin{array}{ccc}
    0.1 & 0.9\\
    0.9 & 0.1\\
  \end{array}
\right]$ and $\left[
\begin{array}{ccc}
    0.25 & 0.75\\
    0.75 & 0.25\\
  \end{array}
\right]$ , respectively. Note that
channel settings 1, 2, 3, and 4 are \textit{non-negatively correlated},
while channel settings 5, 6, and 7 are \textit{negatively correlated}. In
addition, we fix the channel/server number to 10, i.e., $n=10$. 
\begin{figure}[htbp]
\centering
\includegraphics[width=3.2in]{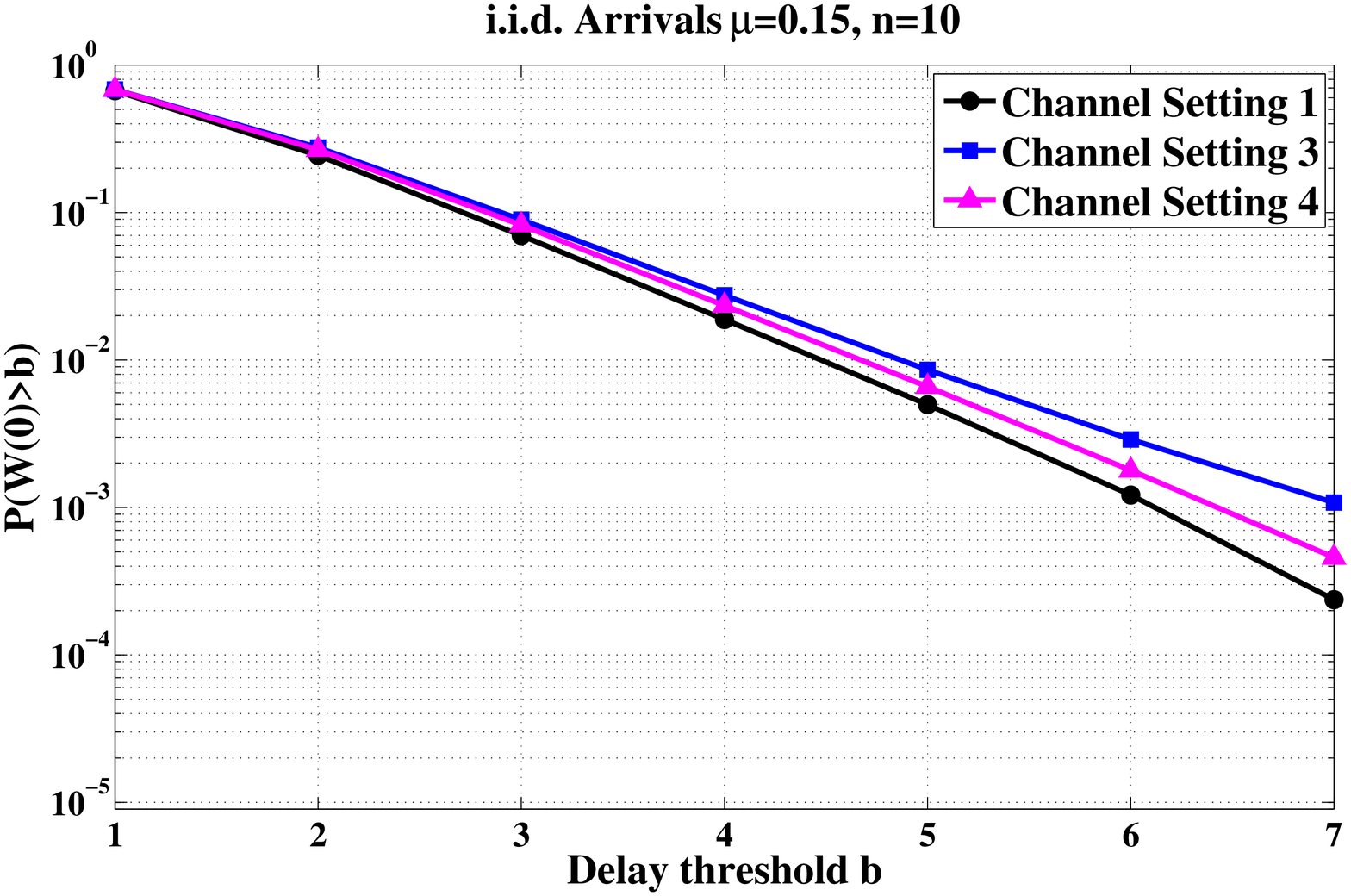}
\caption{Performance comparison under different channel settings with $\mu=0.15$, $n=10$. Channels are \textit{i.i.d.} in channel setting 1 and from channel setting 4 to 3, channels become more \textit{positively correlated}.}\label{fig:delay1}
\end{figure}

\begin{figure}[htbp]
\centering
\includegraphics[width=3.2in]{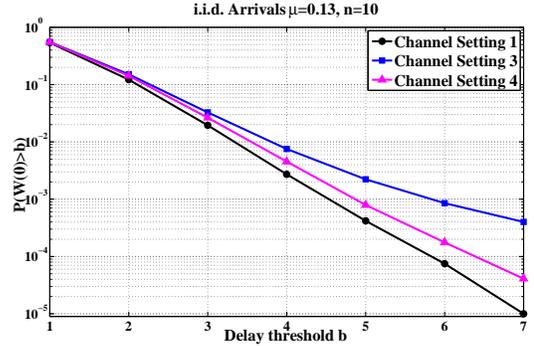}
\caption{Performance comparison under different channel settings, with $\mu=0.13$, $n=10$. Channels are \textit{i.i.d.} in channel setting 1 and from channel setting 4 to 3, channels become more \textit{positively correlated}.}\label{fig:delay2}
\end{figure}
First, we plot the delay violation probability against different
delay thresholds $b$ under channel settings 1, 3, 4 for $\mu=0.15$ and $\mu=0.13$, respectively. From Fig. \ref{fig:delay1} and Fig. \ref{fig:delay2}, we can observe that the \textit{positively correlated} Markovian
channel settings have a larger delay than that in the \emph{i.i.d.} channel setting. This result can also be seen through our
theoretical results. The \emph{i.i.d.} channel setting has a larger delay
rate-function which implies good delay performance. Also,
we can use a single-queue single-server system to mimic the
multi-queue multi-server system here. As channels are more
\textit{positively correlated}, it is more likely to see longer ``ON"
and ``OFF" periods. In this case, the sum of the total service
rate could be very large (up to $n$) or very small with a
non-trivial probability. However, in the \emph{i.i.d.} case, according
to the Chernoff bound, the sum of total service rate lies in
a neighborhood of the mean value $nq$ with high probability. Thus, the service variation under Markovian channels should
be larger than the counterpart under the \emph{i.i.d.} channels. 

Given the same mean service rate, we know from basic
queueing theory that the Markovian channel setting should
have a larger delay. Moreover, if we further lower the arrival
rate (e.g., decrease $\mu$ from 0.15 to 0.13), the simulation
results show that as the channels become more \textit{positively
correlated} the delay gap increases further.
\begin{figure}[htbp]
\centering
\includegraphics[width=3.2in]{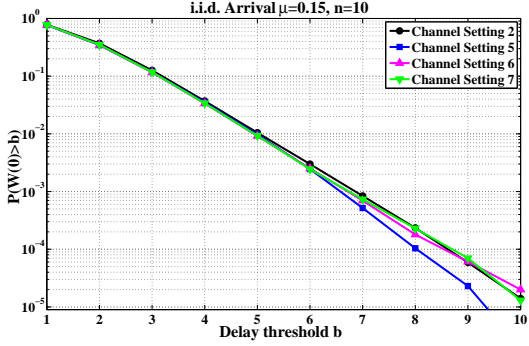}
\caption{Performance comparison under different channel settings, with $\mu=0.15$, $n=10$. Channels are \textit{i.i.d.} in channel setting 2 and from channel setting 7 down to 5, channels become more \textit{negatively correlated}}\label{fig:delay3}
\end{figure} 

Next, we would like to explore the story under negatively
correlated channels. As before, we plot the delay violation
probability against different $b$ under channel settings 2, 5,
6, and 7 for $\mu=0.15$. 
As we can see from Fig. \ref{fig:delay3}, when
channel becomes more \textit{negatively correlated}, the system has
a smaller delay.
An extreme example is channel setting 5,
where alternating ON-OFF-ON... will be observed with high
probability. Once the initial state is determined, the service rate of the system is almost deterministic.
According to basic queueing theory, smaller service variation should give us a
smaller delay. However, when we look at channel setting 6
and 7, there is no big difference between itself and the \emph{i.i.d.} channel setting. 
Therefore, there is still some space for us to find a better scheduling policy 
under the \textit{negatively correlated} channel model.

\section{Conclusion}\label{Conclusion}
In this paper, we considered the scheduling problem of an OFDM downlink system with multiple users and multiple sub-carriers with time-correlated channels. Our theoretical result shows that the class of \textit{oldest packets first} (OPF) policies, which give a higher priority to large delay packets, is delay rate-function optimal when two conditions are both satisfied: 1) The channel is \textit{non-negatively correlated}, and 2) The distribution of ``OFF" period has a memoryless property. 
An open problem for future work is to consider multi-rate channels rather than ON/OFF channels with a unit capacity. 
In this multi-rate channel model, a lexicographically-optimal algorithm that makes the HOL delays most balanced over all the queues is expected to achieve good delay performance. However, the channel-rate heterogeneity introduces a new trade-off between maximizing instantaneous throughput and balancing delays. 
Nonetheless, we believe that the results in this paper will provide useful insights for designing high-performance scheduling policies for more general scenarios.

\appendices
\section{Proof of Lemma \ref{Thm:perfect_matching}}\label{app_X_F}
Applying the law of total probability to different values of $C_{i, j}(t-1)$, we have:
\begin{align}
&\mathbb{P}(C_{i, j}(t)=1|\mathcal{S}(t_1), \mathcal{S}(t_2), \cdots, \mathcal{S}(t_{d}), \mathcal{S}(t-1))\nonumber\\
&=\mathbb{P}(C_{i, j}(t)=1|\mathcal{S}(t_1), \cdots, \mathcal{S}(t_{d}), C_{i, j}(t-1)=1)\nonumber\\
&\qquad\cdot\mathbb{P}(C_{i, j}(t-1)=1|\mathcal{S}(t_1), \cdots, \mathcal{S}(t_{d}), \mathcal{S}(t-1))\nonumber\\
&+\mathbb{P}(C_{i, j}(t)=1|\mathcal{S}(t_1), \cdots, \mathcal{S}(t_{d}), C_{i, j}(t-1)=0)\nonumber\\
&\qquad\cdot\mathbb{P}(C_{i, j}(t-1)=0|\mathcal{S}(t_1), \cdots, \mathcal{S}(t_{d}), \mathcal{S}(t-1)). \label{lemma_first}
\end{align}

Recall that $\hat{T}(t, C_{i, j})$ is the length of the time period from the beginning of its last state-change (``ON" to ``OFF" or ``OFF" to ``ON") time-slot before time-slot $t$ to the beginning of time-slot $t-1$. Summing up all possible values for $\hat{T}(t, C_{i, j})$, we have:
\begin{align}
&\mathbb{P}(C_{i, j}(t)=1|\mathcal{S}(t_1), \mathcal{S}(t_2), \cdots, \mathcal{S}(t_{d}), C_{i, j}(t-1)=1)\nonumber\\
&=\sum_{k=0}^{\infty}\Big{(}\mathbb{P}(C_{i, j}(t)=1|\mathcal{S}(t_1), \cdots, \mathcal{S}(t_{d}), C_{i, j}(t-1)=1, \nonumber\\
&\hat{T}(t, C_{i, j})=k)\nonumber\\
&\times\mathbb{P}(\hat{T}(t, C_{i, j})=k|\mathcal{S}(t_1), \cdots, \mathcal{S}(t_{d}), C_{i, j}(t-1)=1)\Big{)}.
\end{align}

Note that $C_{i, j}(t)$ only depends on the last known state (here is $C_{i, j}(t-1)$) and the last state-change time-slot before time-slot $t$, thus, we can simplify the above equation as:
\begin{align}
&\mathbb{P}(C_{i, j}(t)=1|\mathcal{S}(t_1), \mathcal{S}(t_2), \cdots, \mathcal{S}(t_{d}), C_{i, j}(t-1)=1)\nonumber\\
&=\sum_{k=0}^{\infty}\Big{(}\mathbb{P}(C_{i, j}(t)=1|C_{i, j}(t-1)=1, \hat{T}(t, C_{i, j})=k)\nonumber\\
&\times\mathbb{P}(\hat{T}(t, C_{i, j})=k|\mathcal{S}(t_1), \cdots, \mathcal{S}(t_{d}), C_{i, j}(t-1)=1)\Big{)}\nonumber\\
&=\sum_{k=0}^{\infty}\frac{\mathbb{P}(U\geq k+2)}{\mathbb{P}(U\geq k+1)}\nonumber\\
&\cdot\mathbb{P}(\hat{T}(t, C_{i, j})=k|\mathcal{S}(t_1), \cdots, \mathcal{S}(t_{d}), C_{i, j}(t-1)=1)\nonumber\\
&\geq\min_{k\in\{0, 1, \cdots \}}\frac{1-F_U(k+1)}{1-F_U(k)}. \label{case:C=1}
\end{align}

Applying the same method, we have:
\begin{align}
\mathbb{P}&(C_{i, j}(t)=1|\mathcal{S}(t_1), \mathcal{S}(t_2), \cdots, \mathcal{S}(t_{d}), C_{i, j}(t-1)=0)\nonumber\\
&\geq\min_{k\in\{0, 1, \cdots \}}\frac{\mathbb{P}(D=k+1)}{1-F_D(k)}. \label{case:C=0}
\end{align}

Substitute (\ref{case:C=1}) and (\ref{case:C=0}) into (\ref{lemma_first}),
\begin{align}
&\mathbb{P}(C_{i, j}(t)=1|\mathcal{S}(t_1), \mathcal{S}(t_2), \cdots, \mathcal{S}(t_{d}), \mathcal{S}(t-1))\nonumber\\
&\geq \min_{k\in\{0, 1, \cdots \}}\frac{1-F_U(k+1)}{1-F_U(k)}\nonumber\\
&\cdot\mathbb{P}(C_{i, j}(t-1)=1|\mathcal{S}(t_1), \cdots, \mathcal{S}(t_{d}), \mathcal{S}(t-1))\nonumber\\
&+\min_{k\in\{0, 1, \cdots\}}\frac{\mathbb{P}(D=k+1)}{1-F_D(k)}\nonumber\\
&\cdot\mathbb{P}(C_{i, j}(t-1)=0|\mathcal{S}(t_1), \cdots, \mathcal{S}(t_{d}), \mathcal{S}(t-1))\nonumber\\
&=\min\{\min_{k\in\{0, 1, \cdots\}}\frac{1-F_U(k+1)}{1-F_U(k)}, \min_{k\in\{0, 1, \cdots\}}\frac{\mathbb{P}(D=k+1)}{1-F_D(k)}\}\nonumber\\
&=\hat{q}.
\end{align}

The above result gives us the lower bound on the conditional probability that $C_{i, j}(t)=1$ given $\mathcal{S}(t_1), \cdots, \mathcal{S}(t_{d}), \mathcal{S}(t-1)$, thus, by simply replacing $q$ with $\hat{q}$ in the proof of Lemma 6 in \cite{Lin_Tech_Report}, the result stated in the lemma follows.

\section{Proof of Lemma \ref{lem_sum}}\label{app_sum}
From Lemma \ref{Thm:perfect_matching}, there exists an $N_F>0$, for all $n>N_F$, the probability that $X_F(t)=0$ occurs given the connectivity at time-slots $t_1, \cdots, t_d, t-1$ can be bounded as,
\begin{align}
\mathbb{P}&(X_F(t)=0|\mathcal{S}(t_1), \cdots, \mathcal{S}(t_{d}), \mathcal{S}(t-1))\nonumber\\
&\leq\big{(}\frac{n}{1-\hat{q}}\big{)}^{7H}e^{-n\log\frac{1}{1-\hat{q}}}. \label{inequ:F_X=0}
\end{align}

Now, we are seeking an upper bound on the probability that there are exactly $t+a$ time-slots that satisfy $X_F(t)=1$ among all $t+b$ time-slots during the time interval $[-t-b, -1]$.
\begin{align}
&\mathbb{P}(\sum_{\tau=-t-b}^{-1}X_F(\tau)=t+a)\nonumber\\
&\leq\mathbb{P}\Big{(}\bigcup_{t_1<t_2<\cdots<t_{b-a}}X_F(t_1)=0, \cdots,X_F(t_{b-a})=0\Big{)}\nonumber\\
&\leq{t+b \choose t+a} \max_{t_1<t_2<\cdots<t_{b-a}}\mathbb{P}\Big{(}X_F(t_1)=0,\cdots, X_F(t_{b-a})=0\Big{)}. \label{inequ:t+b_t+a}
\end{align}

Applying the chain rule of conditional probability, we have:
\begin{align}\label{equ:chain_rule}
&\mathbb{P}\Big{(}X_F(t_1)=0,\cdots, X_F(t_{b-a})=0\Big{)}\nonumber\\
&=\mathbb{P}(X_F(t_1)=0)\mathbb{P}(X_F(t_2)=0|X_F(t_1)=0)\times\cdots\nonumber\\
&\times\mathbb{P}(X_F(t_{b-a})=0|X_F(t_1)=0,\cdots, X_F(t_{b-a-1})=0).
\end{align}

Next, we consider the R. H. S. of (\ref{equ:chain_rule}). The upper bound on the first term is quite obvious: substituting $q$ by the stationary probability $1-\pi_0$ in lemma 6 in \cite{Lin_Tech_Report}, we have:
\begin{align}
\mathbb{P}(X_F(t_1)=0)\leq\big{(}\frac{n}{\pi_0}\big{)}^{7H}e^{n\log\pi_0}. \label{first_unconditionally}
\end{align}

For the $d^{th}(d>1)$ term, it is the probability of $\{X_F(t_{d})=0\}$ happens given $\{X_F(t_1)=0,\cdots, X_F(t_{d-1})=0\}$ occurs. Now, we want to obtain a bound for $\mathbb{P}(X_F(t_{d})=0|X_F(t_1)=0, \cdots, X_F(t_{d-1})=0)$:
\begin{align}
&\mathbb{P}(X_F(t_{d})=0|X_F(t_1)=0, \cdots, X_F(t_{d-1})=0)\nonumber\\
&=\mathbb{P}(X_F(t_{d})=0|X_F(t_1)=0,\cdots,\nonumber\\
&\qquad\qquad\qquad X_F(t_{d-1})=0, X_F(t_d-1)=0)\nonumber\\
&\times\mathbb{P}(X_F(t_d-1)=0|X_F(t_1)=0,\cdots, X_F(t_{d-1})=0)\nonumber\\
&+\mathbb{P}(X_F(t_{d})=0|X_F(t_1)=0, \cdots, \nonumber\\
&\qquad\qquad\qquad X_F(t_{d-1})=0, X_F(t_d-1)=1)\nonumber\\
&\times\mathbb{P}(X_F(t_d-1)=1|X_F(t_1)=0, \cdots, X_F(t_{d-1})=0).
\end{align}

We use $\mathbf{S}$ to represent the connectivity $\mathcal{S}(\cdot)$ at each time-slot, and define $\mathcal{F}$ to be a collection of all possible vectors $\mathbf{S}$ such that $X_F(t_1)=0,\cdots,X_F(t_{d-1})=0, X_F(t_d-1)=0$. Then we evaluate the following term:
\begin{align}
&\mathbb{P}(X_F(t_{d})=0|X_F(t_1)=0,\cdots,\nonumber\\
&\qquad \qquad\qquad X_F(t_{d-1})=0, X_F(t_d-1)=0)\nonumber\\
&=\sum_{\mathbf{S}\in\mathcal{F}}\mathbb{P}(X_F(t_{d})=0|\mathbf{S})\cdot\mathbb{P}(\mathbf{S}|\mathcal{F})\nonumber\\
&=\sum_{\mathbf{S}\in\mathcal{F}}\mathbb{P}(X_F(t_{d})=0|\mathcal{S}(t_1), \cdots, \mathcal{S}(t_{d-1}), \mathcal{S}(t_d-1))\cdot\mathbb{P}(\mathbf{S}|\mathcal{F})\nonumber\\
&\leq\big{(}\frac{n}{1-\hat{q}}\big{)}^{7H}e^{-n\log\frac{1}{1-\hat{q}}}\sum_{\mathbf{S}\in\mathcal{F}}\mathbb{P}(\mathbf{S}|\mathcal{F})\nonumber\\
&=\big{(}\frac{n}{1-\hat{q}}\big{)}^{7H}e^{-n\log\frac{1}{1-\hat{q}}}.
\end{align}
where the inequality comes from Lemma \ref{Thm:perfect_matching}. Similarly, we have the same result for $\mathbb{P}(X_F(t_{d})=0|X_F(t_1)=0,\cdots,X_F(t_{d-1})=0, X_F(t_d-1)=1)$, thus we have
\begin{align}
&\mathbb{P}(X_F(t_{d})=0|X_F(t_1)=0, \cdots, X_F(t_{d-1})=0)\nonumber\\
&\leq\big{(}\frac{n}{1-\hat{q}}\big{)}^{7H}e^{-n\log\frac{1}{1-\hat{q}}}. \label{conditional_bounds}
\end{align}

Combining what we have already derived in (\ref{first_unconditionally}) and (\ref{conditional_bounds}), we have:
\begin{align}
&\mathbb{P}\Big{(}X_F(t_1)=0,\cdots, X_F(t_{b-a})=0\Big{)}\nonumber\\
&\leq\big{(}\frac{n}{\pi_0}\big{)}^{7H}\big{(}\frac{n}{1-\hat{q}}\big{)}^{7bH}e^{-n\{-\log\pi_0+(b-a-1)\log\frac{1}{1-\hat{q}}\}}.
\end{align}

This inequality holds for any $t_1, t_2, \cdots, t_{b-a}$, hence,
\begin{align}
&\max_{t_1,\cdots, t_{b-a}}\mathbb{P}\Big{(}X_F(t_1)=0,\cdots, X_F(t_{b-a})=0\Big{)}\nonumber\\
&\leq\big{(}\frac{n}{\pi_0}\big{)}^{7H}\big{(}\frac{n}{1-\hat{q}}\big{)}^{7bH}e^{-n\{-\log\pi_0+(b-a-1)\log\frac{1}{1-\hat{q}}\}}.
\end{align} 

Thus, we have for all $a\leq b-1$:
\begin{align}
&\mathbb{P}(\sum_{\tau=-t-b}^{-1}X_F(\tau)=t+a)\nonumber\\
&\leq{t+b \choose t+a} \big{(}n\cdot\frac{1}{\pi_0}\big{)}^{7H}\big{(}\frac{n}{1-\hat{q}}\big{)}^{7bH}e^{-n\{-\log\pi_0+(b-a-1)\log\frac{1}{1-\hat{q}}\}}\nonumber\\
&\leq 2^{t+b}\big{(}\frac{n}{\pi_0}\big{)}^{7H}\big{(}\frac{n}{1-\hat{q}}\big{)}^{7bH}e^{-n\{-\log\pi_0+(b-a-1)\log\frac{1}{1-\hat{q}}\}}.
\end{align}


\section*{Acknowledgment}
This work is funded in part by NSF grants CNS-1446582, CNS-1421576, and CNS-1518829, ONR grant N00014-15-1-2166, and ARO grant W911NF-14-1-0368. 




%


\end{document}